\documentclass[twocolumn,english,aps,pre,amsmath,amssymb,superscriptaddress,longbibliography]{revtex4-1}

\usepackage{algorithm}
\usepackage{amsmath}
\usepackage{amssymb}
\usepackage{amsfonts}
\usepackage[noend]{algpseudocode}
\usepackage{braket}
\usepackage{braket,slashed}
\usepackage{color}
\usepackage{dsfont}
\usepackage[mathscr]{euscript}
\usepackage{xfrac}
\usepackage{bm}
\usepackage{enumitem}
\usepackage{float}
\usepackage{framed}
\usepackage{graphicx}
\usepackage{graphics}
\usepackage[colorlinks=true, breaklinks=true, linkcolor=red, citecolor=blue, urlcolor=blue]{hyperref} 
\usepackage{lipsum}
\usepackage{subfigure}
\usepackage{tikz}
\usepackage[normalem]{ulem}

\allowdisplaybreaks[1]

\definecolor{darkblue}{rgb}{0, 0, 0.8}

\renewcommand{\d}{\ensuremath{\mathrm{d}}}
\newcommand{\e}{\ensuremath{\mathrm{e}}}

\newcommand{\diagram}[1]{\vcenter{\hbox{\includegraphics[scale=0.3,page=#1]{diagrams.pdf}}}}

\begin{document}
\title{Solving frustrated Ising models using tensor networks}
%\title{General construction of numerically contractible tensor networks for frustrated Ising models}
%\title{Tensor network construction for classical frustrated spin systems}
%\title{Resolving Frustration with Coarse Graining}
%title{Mapping Classical Frustration to a Tiling Problem}
\author{Bram Vanhecke}
\email{bavhecke.vanhecke@ugent.be}
\affiliation{Department of Physics and Astronomy, University of Ghent, Krijgslaan 281, 9000 Gent, Belgium}
\author{Jeanne~Colbois}
\email{jeanne.colbois@epfl.ch}
\affiliation{Institute of Physics, Ecole Polytechnique F\'{e}d\'{e}rale de Lausanne (EPFL), CH-1015 Lausanne,~Switzerland}
\author{Laurens~Vanderstraeten}
\affiliation{Department of Physics and Astronomy, University of Ghent, Krijgslaan 281, 9000 Gent, Belgium}
\author{Frank~Verstraete}
\affiliation{Department of Physics and Astronomy, University of Ghent, Krijgslaan 281, 9000 Gent, Belgium}
\author{Fr\'{e}d\'{e}ric~Mila}
\affiliation{Institute of Physics, Ecole Polytechnique F\'{e}d\'{e}rale de Lausanne (EPFL), CH-1015 Lausanne,~Switzerland}

\begin{abstract}
Motivated by the recent success of tensor networks to calculate the residual entropy of spin ice and kagome Ising models, we develop a general framework to study frustrated Ising models in terms of infinite tensor networks %, i.e. tensor networks 
that can be contracted using standard algorithms for infinite systems. This is achieved by reformulating the problem as local rules for configurations on  overlapping clusters chosen in such a way that they relieve the frustration, i.e. that the energy can be minimized independently on each cluster. We show that optimizing the choice of clusters, including the weight on shared bonds, is crucial for the contractibility of the tensor networks, and we derive some basic rules and a linear program to implement them. We illustrate the power of the method by computing the residual entropy of a frustrated Ising spin system on the kagome lattice with next-next-nearest neighbour interactions, vastly outperforming Monte Carlo methods in speed and accuracy. The extension to finite-temperature is briefly discussed.
%Classical frustrated spin systems give rise to many fascinating many-body phenomena, but standard computational techniques have great difficulties in simulating a large variety of them. While tensor networks have obtained promising results in a number of cases, we show that the standard tensor network construction for classical partition functions cannot be used in general. Instead, we argue that for frustrated systems, successful tensor network formulations deal with frustration at the level of the tensor. This is achieved by reformulating the classical statistical mechanics problem into a frustration-free tiling problem, resulting both in an improved analytical understanding of the ground state ensembles via local rules, and a natural construction of a tensor network enabling high precision computations. An essential insight is that the tiles have to be selected with care for the tensor network methods to converge. The main technical ingredient therefore consists in a linear program systematically identifying the large scale degrees of freedom which are used as tiles. We illustrate the power of the method by determining the ground state local rule and computing the residual entropy of a frustrated Ising spin system on the kagome lattice with next-next-nearest neighbour interactions, vastly outperforming Monte Carlo methods in speed and accuracy.
\end{abstract}

\date{\today}
\maketitle

\renewcommand{\diagram}[1]{\;\vcenter{\hbox{\includegraphics[scale=0.32,page=#1]{./diagrams.pdf}}}\;}

\section{Introduction}
One of the most beautiful manifestations of emergent behaviour in statistical physics can be found in the arena of frustrated spin systems \cite{Lacroix2011}. Frustration in a classical spin system occurs whenever it is impossible to find a spin configuration which minimises each and every term of the Hamiltonian simultaneously, leading to macroscopic ground state degeneracies and giving rise to interesting zero-temperature physics such as effective realizations of gauge theories \cite{Henley2010}.
\par Early exact results in this context were obtained for antiferromagnetic nearest-neighbour Ising models on the triangular and kagome lattices \cite{Wannier1950,Kano1953} using Kauffman and Onsager's method; for frustrated Ising models on all planar two-dimensional lattices with nearest-neighbour interactions using a mapping to free fermions \cite{Kasteleyn1963, Schultz1964}; and for more general systems such as planar spin ice \cite{Lieb1967} using Bethe ansatz techniques \cite{Baxter1982}. 
\par It has, however, proven difficult to treat frustration in generic (i.e., non-integrable) models: to reach the low-energy phase space and sample it efficiently, Monte Carlo methods require ad-hoc non-local cluster updates to fight both critical slowing down \cite{Swendsen1987, Wolff1989} and  frustration \cite{Wang2012, Rakala2017}. In addition, calculating the free energy requires the use of thermodynamic integration, making zero-temperature residual entropies hard to determine accurately.
\par Tensor networks \cite{Verstraete2008} provide a new computational approach for studying ground states of classical lattice models with strong correlations, as was recently demonstrated by the determination of the residual entropy of ice and dimer models in three-dimensional lattices with unprecedented precision \cite{Vanderstraeten2018}. 
%This was achieved by employing matrix product state (MPS) and projected entangled-pair state (PEPS) algorithms to determine the leading eigenvectors of row-to-row or plane-to-plane transfer matrices. 
There is some freedom when expressing a partition function as a tensor network. The formulation of the tensor network in Ref.~\onlinecite{Vanderstraeten2018} relies on a pre-existing knowledge of ground state \textit{local rules} -- rules that all the ground state configurations of a model must satisfy -- easily implemented at the level of the tensor.
\par In this paper, we generalise the applicability of tensor networks to more generic frustrated spin systems. For this, we first revisit nearest-neighbour frustrated systems. We argue that, for a given partition function, the choice of the tensor network expression affects the convergence of the contraction algorithms, and we show that taking the zero temperature limit of the standard formulation \cite{Orus2009,Haegeman2016} is not always an option. We observe that a formulation relying on ground state local rules, turning the computation of the partition function into a tiling problem, appears to be crucial for the contraction to converge. Furthermore, the tiles need to be selected with care, and we illustrate how this can be done. We introduce a linear program (extending on Ref.~\onlinecite{Huang2016}) to systematically look for such local rules in generic further neighbour models, and we show how this yields a natural expression for the tensor network enabling the study of the full ground state manifold. The construction holds the roots of a generalisation to finite temperature. To demonstrate the power of the method, we apply it to a frustrated Ising model with further neighbours couplings on the kagome lattice, obtain the residual entropy with a very high accuracy, and use the local rules to make some exact statements regarding the physics of the ground state manifold.

\section{Standard construction}
\begin{figure}
\centering
\includegraphics[width = 0.4\columnwidth,page=1]{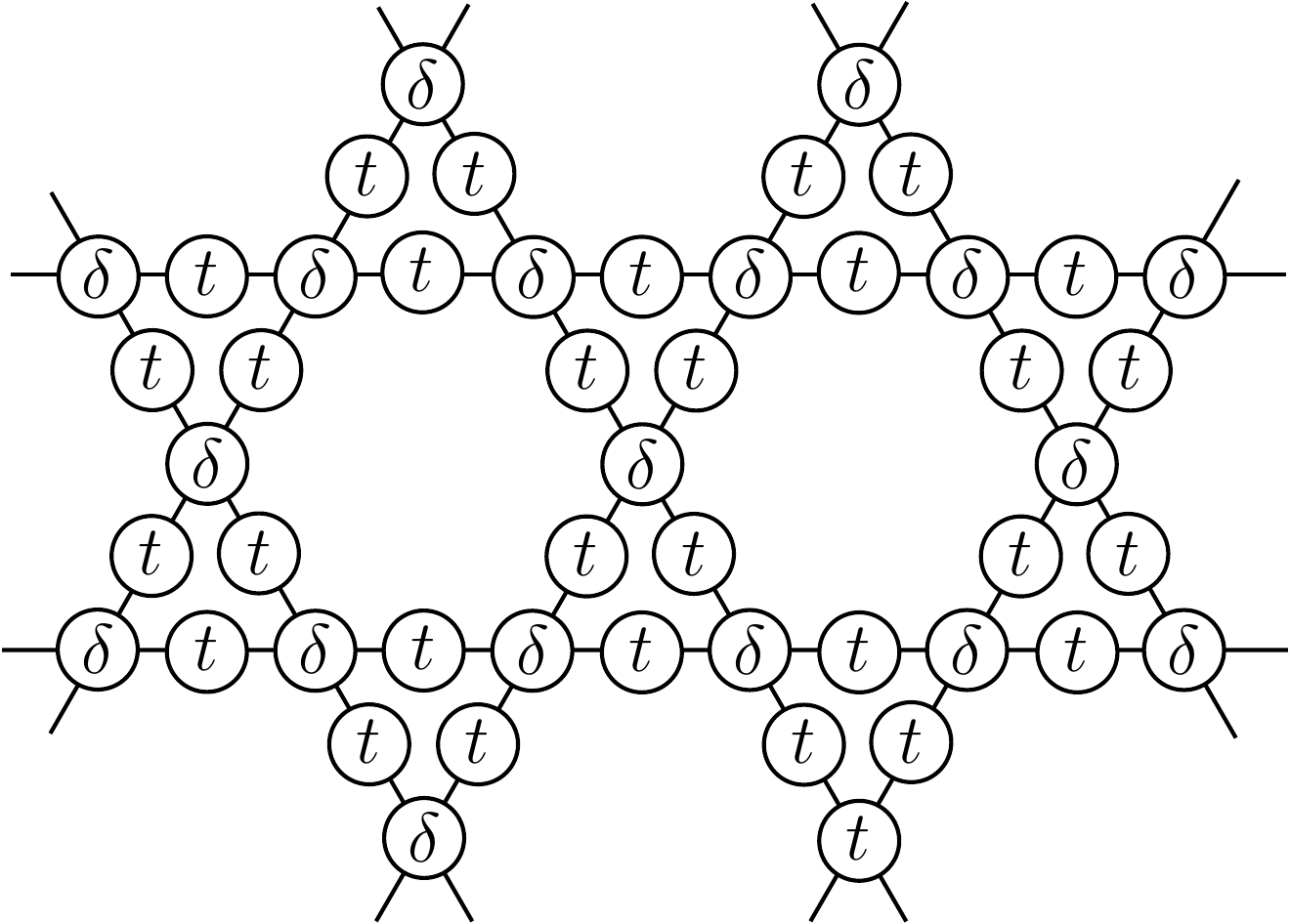} \qquad
\includegraphics[width = 0.4\columnwidth,page=2]{diagrams.pdf}
\caption{Standard tensor network construction for the partition function on the triangular and kagome lattices. The matrices $t$ carry the zmann weights, while the $\delta$ tensors enforce that neighbouring $t$'s share the same spin.}
\label{fig:Z}
\end{figure}
Partition functions for classical spin systems can be expressed as contractions of tensor networks in the spirit of the transfer matrix formalism, a representation which is not unique. The standard construction consists in associating to each interaction a matrix $t$ accounting for its Boltzmann weight, and to each on-site variable a $\delta$ tensor, i.e.~a tensor whose rank corresponds to the number of interactions involving that site and which is 1 only when all its indices take the same value. In the particular case of the kagome and triangular lattice Ising (anti)ferromagnets, this standard formulation leads to the tensor networks of Fig.~\ref{fig:Z}, where the Boltzmann weights are given by
\begin{equation}
\label{eq:boltzmannweight}
t^{\sigma_i, \sigma_j} = \e^{ -\beta J\sigma_i\sigma_j },
\end{equation}
so that the matrix $t$ reads
\begin{equation}
\label{eq:boltzmannweight}
t=\begin{pmatrix} \e^{-\beta J} & \e^{+\beta J}  \\ \e^{+\beta J} & \e^{-\beta J} \end{pmatrix} ,
\end{equation}
with $J > 0$ for the antiferromagnet. Contracting the tensor network amounts to finding the leading eigenvalue and leading eigenvector of the row-to-row transfer matrices (see e.g.~\cite{Haegeman2016}). When the algorithm converges, the logarithm of the leading eigenvalue, directly related to the free energy per site at the given inverse temperature $\beta$, is obtained with an extremely high accuracy. \\
\noindent\textbf{\emph{Issue with the zero temperature limit}}. %
However, it is obvious that low temperature properties cannot be directly probed from the standard construction since the zero temperature limit of Eq.~\eqref{eq:boltzmannweight} cannot be taken. Similarly, partition function is ill-defined in that limit. In simple cases, this problem can be solved by considering the regularized partition function $\mathcal{Z}_0$ whose zero temperature limit is always well defined and is directly related to the residual entropy:
\begin{equation}
\label{eq:residualentropy}
	\mathcal{Z}_0	:=e^{\beta E_0 N} \mathcal{Z} \qquad S = \lim_{\beta\rightarrow \infty} \frac{1}{N} \ln(\mathcal{Z}_0)
\end{equation}
where we have introduced the ground state energy per site $E_0$ and the number of sites $N$.
\par Indeed, to compute $\mathcal{Z}_0$ instead of $\mathcal{Z}$, one has to contract the tensor network based on
$t_{0}^{\sigma_i,\sigma_j} = e^{\beta \frac{E_0}{z} }t^{\sigma_i,\sigma_j}$ where $z$ is the number of bonds per site.
In a nonfrustrated system, all the pair interactions are minimised simultaneously, removing all exponentially diverging matrix elements, hence ensuring that the zero temperature limit can be taken both in $t_0$ and $\mathcal{Z}_0$. However, in a frustrated system, the pair interactions cannot be minimised simultaneously and $t_0$ still contains exponentially diverging factors. For instance, compare the tensors for the ferromagnetic and antiferromagnetic Ising models on the kagome lattice \footnote{The tensors are the same for the triangular lattice Ising antiferromagnet.}: 
\begin{equation}
	t_0^{\text{F}} = \begin{pmatrix} 1 & \e^{-2\beta |J|}  \\ \e^{-2\beta |J|} & 1 \end{pmatrix} \quad t_0^{\text{AF}} = \begin{pmatrix} \e^{- \frac{4}{3} \beta J} & \e^{\frac{2}{3}\beta J}  \\ \e^{\frac{2}{3}\beta J} & \e^{-\frac{4}{3}\beta J} \end{pmatrix} .
\end{equation}
%\begin{figure}
%\centering
%\includegraphics[width = \columnwidth]{KagomeTN.png}
%\caption{\label{fig:kagomeTN}Tensor network construction for the ground state of the kagome lattice Ising antiferromagnet: the Hamiltonian can be split up into triangular terms. The configurations which minimise the triangular Hamiltonian can be tiled to create ground states. We can build a tensor network to count the tilings and study the ground state manifold by associating a tensor to each triangle, and introducing bond matrices $P$ enforcing that spins must match. The bond dimension is significantly reduced by performing an SVD on $P$. At finite temperature, the construction is the same, only the tensors are promoted to consider more configurations and associate them a Boltzmann weight.}
%\end{figure}
\renewcommand{\diagram}[1]{\;\vcenter{\hbox{\includegraphics[scale=0.27,page=#1]{./diagrams.pdf}}}\;}
\begin{figure}
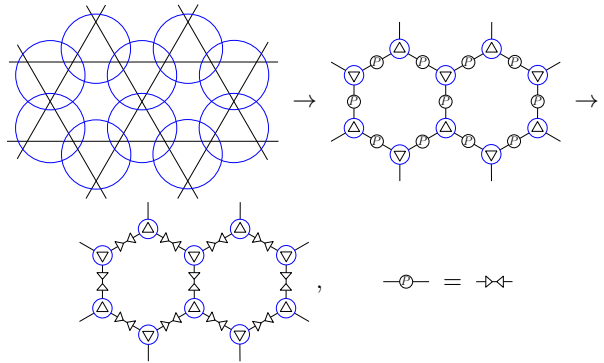

\centering
$\diagram{3}\to \diagram{4} \to \diagram{5}, \qquad \diagram{6} = \diagram{7}$
\caption{\label{fig:kagomeTN}Tensor network construction for the ground state of the kagome lattice Ising antiferromagnet: the Hamiltonian can be split up into triangular terms. The configurations which minimise the triangular Hamiltonian can be tiled to create ground states. We can build a tensor network to count the tilings and study the ground state manifold by associating a tensor to each triangle, and introducing bond matrices $P$ enforcing that spins must match. The bond dimension is significantly reduced by performing an SVD on $P$. At finite temperature, the construction is the same, only the tensors are promoted to consider more configurations and provide them with a Boltzmann weight.}
\end{figure}

\section{Nearest-neighbour Ising antiferromagnet on the kagome lattice}
\noindent\textbf{\emph{Ground state}}. %
To build up to a generic construction, we start by re-exploring the simple frustrated models where a solution is known. The simplest is the kagome lattice Ising antiferromagnet. In this model, the ground state configurations have to satisfy a ``2-up 1-down, 2-down 1-up'' rule (no ferromagnetic triangles). This is easily seen by writing the Hamiltonian as a sum of triangular Hamiltonians
\begin{align}
	H = \sum_{\langle i,j \rangle} J \sigma_i \sigma_j = \sum_{\substack{\vartriangle_{i,j,k}\\ \triangledown_{i,j,k}}} J(\sigma_i \sigma_j + \sigma_j \sigma_k + \sigma_k \sigma_i)\\
	:=  \sum_{\vartriangle_{i,j,k}} H_{\vartriangle_{i,j,k}} + \sum_{\triangledown_{i,j,k}} H_{\triangledown_{i,j,k}}.
\end{align}
One triangular Hamiltonian is minimised by non-ferromagnetic spin configurations on the triangle. Since the triangular Hamiltonians can be simultaneously minimised, \emph{all} the ground states of the model can be described as tilings of 2-up 1-down, 2-down 1-up triangles on kagome, where the triangular tiles fit if the shared spin is the same.
\par This is easily translated into a tensor network (slightly different from the one in Ref.~\onlinecite{Vanderstraeten2018}) on the (dual) honeycomb lattice. The prescription is as follows (Fig. \ref{fig:kagomeTN}):
\begin{enumerate}
\item On each site of the dual lattice (center of the kagome triangles), place a $\delta$ tensor with rank 3 and bond dimension 6 describing the 6 ground state configurations of this triangle,
\item On each bond of the dual lattice (sites of the kagome lattice), place a bond matrix $P$ with bond dimension 6 which is 1 if the two connected tensor assign the same value to their shared spin, and 0 otherwise,
\item Reduce the bond dimension of the tensor network to 2 by performing a singular value decomposition (SVD) on the $P$ tensors and grouping the resulting tensors with the $\delta$ tensors on the triangles. 
\end{enumerate}
\begin{table*}
	\begin{tabular}{ | p{2cm} | p{3.5cm} | p{3.5cm} |}
		\hline
		& AF-Ising on kagome & AF-Ising on triangular\\ \hline
		MPS & 0.5018331646 ($D=10$) & 0.3230659407 ($D=250$) \\ \hline
		exact & 0.5018331646 & 0.3230659669 \\ \hline
	\end{tabular}
	\caption{Tensor-network results obtained using vumps on the row-to-row transfer matrix ($D$ is the MPS bond dimension). Taken from Ref.~\onlinecite{Vanderstraeten2018}.}
	\label{tbl:ResEntropies}
\end{table*}
This tensor network is well defined and provides the ground state entropy of the kagome lattice with a precision of $10^{-10}$ (Table \ref{tbl:ResEntropies}). This example demonstrates that it can be very useful to use clusters (here triangles) to build a tensor network. However, the choice of clusters, which is rather natural in the case of kagome, is in general a subtle issue, as we now show on the example of the triangular lattice.\\

\begin{figure}
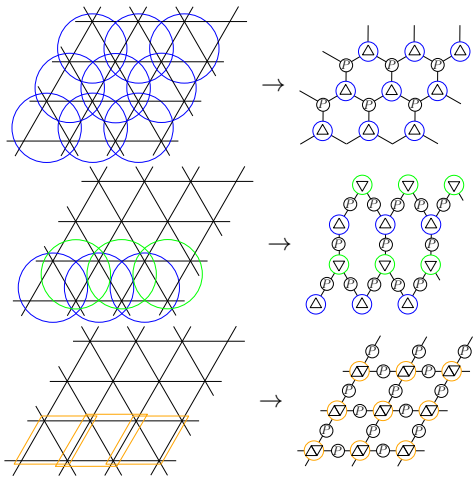

\centering
$\diagram{8}\to \diagram{9} $\\
$\diagram{10}\to \diagram{11} $ \\
$\diagram{12}\to \diagram{13} $
\caption{\label{fig:TLIAFTN}Tensor network construction for the ground state of the triangular lattice Ising antiferromagnet: the Hamiltonian can be split up into triangular terms in various ways. The configurations which minimise the triangular Hamiltonian can be tiled to create ground states. Top: construction corresponding to Eq.~\eqref{eq:HTLIAF1}. Middle: construction corresponding to Eq.~\eqref{eq:HTLIAF2}. Bottom: construction based on the Hamiltonian tessellation Eq.~\eqref{eq:hamtess}. The bond dimension is significantly reduced by performing an SVD on the bond matrices $P$. }
\end{figure}

\section{Nearest-neighbour Ising antiferromagnet on the triangular lattice}

\noindent\textbf{\emph{Ground state}}. %
We proceed with the archetype of frustration: the triangular lattice Ising antiferromagnet. Inspired by the kagome construction, we look for \emph{tiles} that can be used to build \emph{all} the ground states. We want to find them as local ground state configurations of \emph{local} Hamiltonians (the equivalent of the triangular Hamiltonian in the previous section) which can be \emph{simultaneously minimised}. This last criterion is essential to ensure that \emph{each} ground state can be described using these tiles.\\
For this, we notice that the Hamiltonian can be written as a sum of terms acting only on one type of triangles, for instance $\vartriangle$ triangles:
\begin{equation}
	\label{eq:HTLIAF1}	H = \sum_{\langle i,j \rangle} J \sigma_i \sigma_j = \sum_{\vartriangle_{i,j,k}} J(\sigma_i \sigma_j + \sigma_j \sigma_k + \sigma_k \sigma_i).
\end{equation}
The triangular Hamiltonian is minimised by non-ferromagnetic triangles. Since there exists a state which minimises all the triangular Hamiltonians, the set of all ground states can be obtained by tiling non-ferromagnetic down triangles, which fit if the spins in the overlap of three triangles have the same orientation \footnote{Indeed, there cannot be a ground state containing a ferromagnetic down triangle: this would mean that one of the triangular Hamiltonians is not minimised}. The corresponding tensor network has $\delta$ tensors at the centers of up triangles and rank-3 $P$ tensors enforcing the consistency on the spin shared by three $\delta$ tensors (Fig. \ref{fig:TLIAFTN}).
\par Another valid splitting of the Hamiltonian is to share bonds between up and down triangles:
\begin{equation}
	\label{eq:HTLIAF2}
	H = \sum_{\langle i,j \rangle} J \sigma_i \sigma_j = \sum_{\substack{\vartriangle_{i,j,k}\\ \triangledown_{i,j,k}}} \frac{J}{2}(\sigma_i \sigma_j + \sigma_j \sigma_k + \sigma_k \sigma_i).
\end{equation}
This splitting amounts to tiling non-ferromagnetic up \textit{and} down triangles with the condition that on a shared bond, the two spins must match; the corresponding tensor network is defined on the honeycomb lattice and has $\delta$ tensors on up \textit{and} on down triangles, with bond matrices $P$ now taking care of two spins (Fig. \ref{fig:TLIAFTN}). Note that we have chosen to give the tensor network minimal connectivity: not all triangles that share a spin are connected, some shared spins are implicitly enforced to be the same via multiple bonds.
\par
\begin{figure}
	\includegraphics[width = \columnwidth]{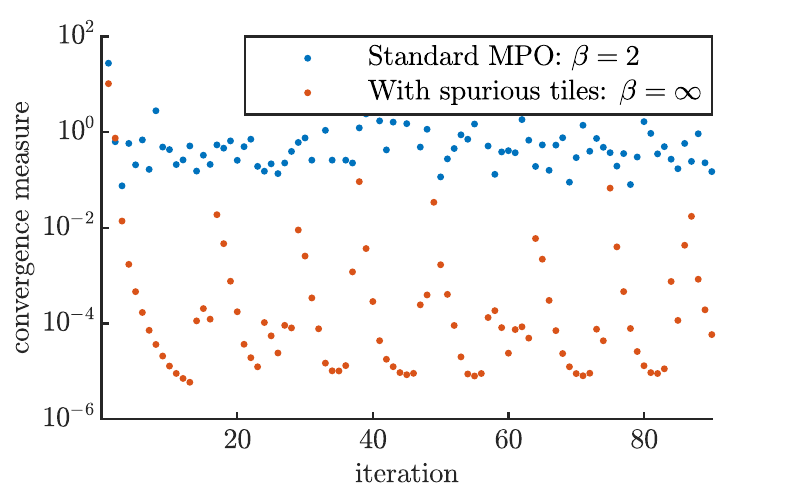}\\
	\includegraphics[width = \columnwidth]{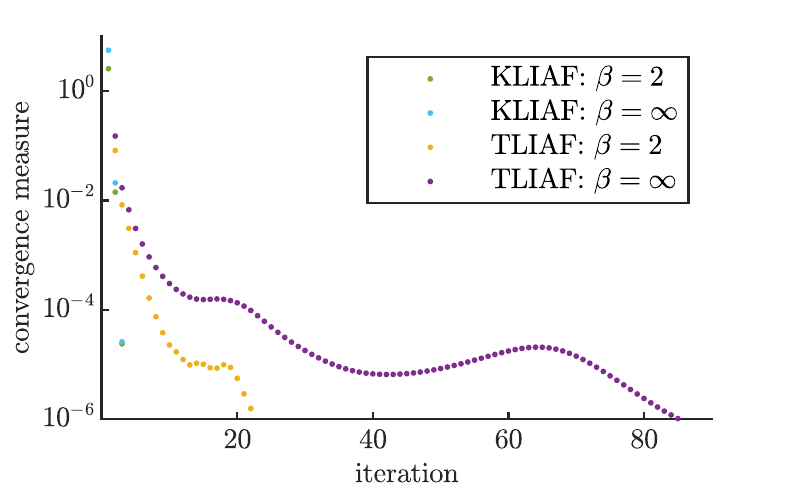}
	\caption{Top: Convergence of vumps algorithm \cite{ZaunerStauber2018, Fishman2018, Vanderstraeten2019} for the antiferromagnetic triangular lattice Ising model at $\beta=2$ using the standard tensor network from Fig. \ref{fig:Z} and at $\beta=\infty$ using the construction based on Eq.~\eqref{eq:HTLIAF1}. Bottom: vumps convergence for the antiferromagnetic Kagome lattice Ising model at $\beta=2$ and $\beta=\infty$, and the same for the antiferromagnetic triangular Ising model using the construction of Eq.~\eqref{eq:hamtess}. Each for MPS with a bond dimension of $\chi=80$ and we use a variational convergence measure, see Ref.~\onlinecite{Vanderstraeten2019}. We observed similar behaviour using the corner transfer matrix renormalization group algorithm \cite{Nishino1996,Orus2009}, and a similar issue was observed for real-space renormalization techniques in Ref.~\onlinecite{Zhu2019}.}
	\label{fig:vumps}
\end{figure}

 These two splittings are equally valid (in the thermodynamic limit or with periodic boundary conditions), and both solve the regularisation problem by working directly in the ground state. However, standard contraction algorithms fail to converge for the first construction, while they converge and lead to the correct answer for the second one (Fig.~\ref{fig:vumps}, Table \ref{tbl:ResEntropies}).
\par The main difference between the two cases is that, while in the second construction the constraint forbidding down triangles to be ferromagnetic is imposed at the level of the tensors, in the first construction it is imposed non-locally. Indeed, since the Hamiltonians in Eqs.~\eqref{eq:HTLIAF1} and~\eqref{eq:HTLIAF2} are the same, they must have the same energy for all global states, implying that if a state contains a down triangle which is ferromagnetic, it must also contain a ferromagnetic up triangle. The key point is that these two triangles can be arbitrarily far apart. Accordingly, approximate algorithms, which are based on large but finite bond dimensions, fail to converge.

%Expecting from an approximate contraction algorithm to correctly assess from the first formulation that a configuration involving a ferromagnetic up triangle must be dismissed is a high requirement.\\
\noindent\textbf{\emph{Hamiltonian tessellation}}. %
In general, we don't have such an insight on the ground state of frustrated models. To generalise tensor network constructions, the question thus boils down to being able to select \textit{a priori} among all possible splittings of the Hamiltonian of interest, the equivalent of Eq.~\eqref{eq:HTLIAF2} and not Eq.~\eqref{eq:HTLIAF1}. Let us see how this can be done for the triangular lattice.
\par These two splittings can be seen as instances of the following generic \emph{Hamiltonian tessellation} (we will use this term to describe a set of ways of splitting the Hamiltonian):
\begin{align} 
\mathcal{H} &= \sum_{c \in \mathcal{T}_u} \sum_{n \in c} \,\alpha^{c}_{n} \,h_{n} = \sum_{c \in \mathcal{T}_u} H^{\{\alpha\}}_c \label{eq:hamtess}\\
& \sum_{c \in \mathcal{T}_u| n \in c} \alpha^{c}_n = 1 \,,\qquad\forall \,n
\label{eq:alpha}
\end{align}
where for later convenience we have considered a cluster $u$ regrouping two triangles:
\begin{equation*}
\diagram{14}.
\end{equation*}
$\mathcal{T}_u$ is the set of clusters obtained from translating $u$ on the lattice (with overlaps). Each cluster is seen as a collection of bonds (indexed by $n$) and to each bond Hamiltonian $h_n$ ($h_{\langle i,j \rangle} = J\sigma_i \sigma_j$) we associate weights $\alpha^c_n$ specifying how much of it is accounted for in each cluster (Eq. \ref{eq:alpha} imposes that terms appearing in a single cluster have weight 1). In the following, to ensure translation invariance, we always choose the $\alpha^c_n$ to be the same for each~$c \in \mathcal{T}_u$ \footnote{In programmatic terms, the Hamiltonian tessellation can be thought of as a class of Hamiltonian splittings described by the choice of the cluster $u$, and where choosing the values for the weights $\{\alpha^c_n\}$ defines an instance of the class}.
In our triangular case, each cluster has five interaction terms, four of which are shared with neighbouring clusters, and the associated weights must satisfy the constraints
\begin{equation}
\label{eq:const1}
\alpha'_{1,2} = 1 - \alpha_{1,2} 
\end{equation}
by translation invariance and Eq.~\eqref{eq:alpha}. 
\par Remember that the aim is to find tiles to build all the ground states. We find these tiles as local ground state configurations  $\{C^{\{\alpha\}}_u\}$  on $u$ minimising the local Hamiltonian $H^{\{\alpha\}}_u$.  Depending on the weights, we get different ground state configurations $\{C^{\{\alpha\}}_u\}$; only the weights for which all the local Hamiltonians can be simultaneously minimised provide tiles which can be used to describe the whole ground state manifold. In our case, this further restricts the weights to 
\begin{equation}
\label{eq:const2}
\alpha_1 = \alpha_2 \in [0,1].
\end{equation}
This is the (convex) set of weights which satisfy
\begin{equation}
H^{\{\alpha\}}(C) \geq -J \, \forall C \text{ on } u
\end{equation}
where $-J$ is the ground state energy per cluster. The boundaries of the convex set are defined by some of these inequalities becoming equalities.\\
In this formulation, we can see in a new light what happens on the triangular lattice. 
By construction, for any weights in the convex set defined by Eqs.~\eqref{eq:const1} and~\eqref{eq:const2}, all the ground states can be constructed as tilings of the local ground state configurations. On the one hand, Eq.~\eqref{eq:HTLIAF2} corresponds to Eq.~\eqref{eq:hamtess} with $\alpha_1 = \alpha_2 = 1/2$. There are 10 local ground state configurations of the unit $u$, hence 10 tiles. These are the configurations containing no ferromagnetic triangles. On the other hand, Eq.~\eqref{eq:HTLIAF1} corresponds to Eq.~\eqref{eq:hamtess} with $\alpha_1 = \alpha_2 = 1$. At this point (which lies on the boundary of the convex set), an additional accidental ground state degeneracy occurs: configurations for which the up triangle is ferromagnetic now have the ground state energy as well. These two additional tiles cannot play a role in the ground state manifold, since there are weights for which they are not ground state tiles; so, they cannot fit into any global ground state. We call such tiles \emph{spurious} because they do not really belong to the ensemble of ground state tiles. Thus, according to our observation that contraction is possible for the tessellation of Eq.~\eqref{eq:HTLIAF2} but not for that of Eq.~\eqref{eq:HTLIAF1}, it sounds like a good idea to get rid of such tiles to ensure the convergence of the tensor network.
\section{Generic implementation}
\begin{figure} \begin{center}
		\includegraphics[page=15,width=0.5\columnwidth]{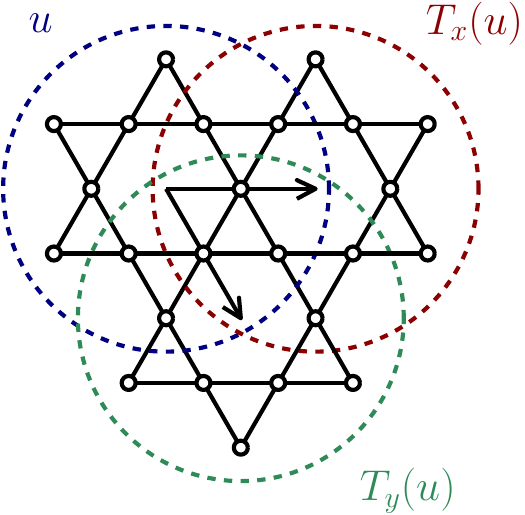}
		\caption{A tessellation of spins on the kagome lattice: the cluster $u$ consists of twelve spins, and it shares five spins with each of the translated clusters $T_x(u)$ and $T_y(u)$.}
		\label{fig:cluster}
	\end{center}
\end{figure}

\noindent\textbf{\emph{Maximal lower bound}}. %
All of the above can be quite straightforwardly adapted for a generic system of $d$-level spins $s_i$ on a lattice, with a translation invariant Hamiltonian $\mathcal{H}$ containing only local interaction terms $h_n$ of strictly bounded range (e.g., finite range further neighbour pair interactions):
\begin{equation}
	H(\{\sigma\}) = \sum_n h_n(\{\sigma\}_n)
\end{equation}
where $\{\sigma\}_n$ denotes the subset of spins taking part in interaction $n$.
Given a reference cluster of spins $u$, we cover the lattice with the set $\mathcal{T}_u$ of overlapping translated $u$'s such that the Hamiltonian can be rewritten as a sum of strictly local terms acting within a single cluster (see for instance  Fig.~\ref{fig:cluster}).
Just like in the triangular case, we associate to each Hamiltonian term $h_n$ weights $\alpha^c_n$ describing how they are shared between clusters, recovering exactly the expression that we gave for the triangular lattice in Eqs.~\eqref{eq:hamtess} and~\eqref{eq:alpha}. Since, in this form, the Hamiltonian contains only terms that act within a cluster, the minimum of $H^{\{\alpha\}}_u$ with respect to the spin configurations of $u$ implies a lower bound on the global ground state energy. This bound can be optimized by maximizing over $\alpha^{u}_{n}$ \cite{Huang2016,Kaburagi75,Kudo76} \footnote{The method described in Ref. \onlinecite{Huang2016} is dual to their ``basic polytope method'' which is itself similar to {Kanamori}'s method developed in Ref. \onlinecite{Kaburagi75}. The main differences are that {Kanamori}'s method has weaker convergence properties but gives lower bounds for all the couplings at once}. This optimisation (which was in particular leveraged in Ref.~\onlinecite{Huang2016} to obtain ground state energies of generalised Ising models) can be expressed as a linear program \cite{Grotschel1988}: 
\begin{equation}
E_u\quad\leftarrow\quad\max_{\vec{\alpha}} \,E\,, \quad \text{with} \quad \left\{ \begin{array}{l}
H^{\{\alpha\}}_u(C)\geq E \; \forall C \\ \sum_{c \in \mathcal{T}_u| n \in c} \alpha^{c}_n = 1 \end{array} \right.,
\label{eq:linprogE}
\end{equation}
where the result of the maximization, $E_u$, is the candidate ground state energy per cluster, and where $C$ goes through the configurations of $u$. This program may be solved using a standard linear programming toolbox. \\
\noindent\textbf{\emph{Getting the most out of knowing the ground state energy}}. %
Our point is that if the maximal lower bound is saturated (i.e. if $u$ is such that there exists a global state which has $E_u$ as energy per cluster, or equivalently when this lower bound matches an upper bound), one gets more than just the ground state energy. Indeed, for the weights $\{\alpha^{u}_n\}$ which realise this maximal lower bound, by construction the local cluster Hamiltonians $H^{\{\alpha\}}_u$ can be simultaneously minimised if and only if the maximal lower bound is saturated. We will say that such a Hamiltonian has \emph{minimal frustration}. In this case, all the ground states are characterised as tilings of the configurations of the cluster $u$ belonging to the set
\begin{equation}
G^{\{\alpha \}} := \{C\,|\,H^{\{\alpha\}}_u(C) = E_u \}.
\end{equation}
In this sense, the set of tiles $G^{\{\alpha \}}$ is a local rule. 
\par There are however many solutions of \eqref{eq:linprogE}, namely all sets of weigths $\{\alpha^{u}_{n}\}$ satisfying
\begin{equation}
H^{\{\alpha\}}_u(C)\geq E_u, \quad \text{for all configurations $C$ of $u$}.
\label{eq:ineq}
\end{equation}
As we have seen in the triangular lattice case, not all $\{\alpha^{u}_{n}\}$ will do. In the space of the weights, the set of $\{\alpha^{u}_{n}\}$ for which all these inequalities are satisfied takes the form of a convex set, which we will refer to as $A^u$, corresponding to the generalisation of Eqs.~\eqref{eq:const2}. The set of ground state tiles $G_u^{\{\alpha \}}$ does not depend on the weights in the \emph{interior} of $A^u$. However, just like in the triangular lattice case, the boundary is defined by some of the inequalities becoming equalities, and accidental degeneracies will occur:
\begin{equation}
G_u^{\{\alpha\}\in \text{Int}(A^u) } \subset G_u^{\{\alpha\} \in \text{Bound}(A^u)}.
\end{equation}
The associated additional configurations must be spurious tiles, which could spoil the contractibility of the tensor network as well as hinder the understanding of the ground state manifold. 
\par Thus, for a generic problem, we need to 
\begin{enumerate}
\item find a cluster $u$ such that the maximal lower bound for the ground state energy, $E_u$, is saturated,
\item find weights $\{\alpha^u_n\}$ in the interior of $A^u$.
\end{enumerate}
Note that the second step allows one to get rid of avoidable spurious tiles, but there could as well be some tiles in $G_u^{\{\alpha\}\in \text{Int}(A^u) }$ which do not belong to any ground state. Getting rid of the avoidable tiles might help, but in general, because of the lack of insight into the problem, several clusters might need to be tested. Additionally, to find a point in the interior of $A^u$, splitting the weights evenly among clusters does not always work. The fact that the problem can be phrased as a linear program is thus very helpful: first, it allows one to rapidly test for various candidate clusters $u$; second, with a bit of additional work as mentioned below, it allows one to enforce the selection of weights in the interior of $A^u$, thus getting rid of avoidable spurious tiles.
\par As a technical note, linear program solvers only output extreme points of the convex set, so simply solving \eqref{eq:linprogE} will systematically give $\{\alpha^u_n\}$ corresponding to avoidable spurious tiles. We show in Appendix~\ref{sec:pseudocode} how to overcome this by finding boundary points that form a simplex of the same effective dimension as $A^u$, ensuring that any point in the interior of this simplex will also lie in the interior of $A^u$ \footnote{A simplex is the generalisation to high dimension of triangles and tetrahedrons.}. Another technical challenge is that the number of constraints scales exponentially in the number of spins per cluster. In Appendix~\ref{sec:pseudocode}, we also show how to work around this problem by systematically and progressively incorporating inequalities as we build the corners of the interior simplex, such that only a very limited number of inequalities is needed.\\
\noindent\textbf{\emph{Generic tensor network}}. %
Finally, the procedure to write a contractible tensor network is easily generalised. For the ground state, the tiles $G_u^{\{\alpha\}\in \text{Int}(A^u) }$ are described by $\delta$ tensors, placed on each dual vertex, coinciding with the clusters of $\mathcal{T}_u$. The overlapping spins matching condition is enforced by bond matrices $P$. %The finite temperature construction is obtained by promoting the $\delta$ tensor to a $D_0$ tensor associating to each configuration on $u$ its Boltzmann weight relative to the ground state energy per cluster.
 Performing an SVD on the rank-deficient bond matrices keeps the tensor network bond dimension reasonably small.\\
\noindent\textbf{\emph{Testing for the saturation of the maximal lower bound}}. %
Provided that the maximal ground state lower bound is saturated, all of the above is given. Rigorously proving that it is the case is equivalent to finding one ground state, or proving that the tiles $G^{\{\alpha\}}_u$ corresponding to the maximal lower bound can tile the plane. In general, this is an undecidable problem \cite{Wang1961, Robinson1971, Wang1975, Wang2017}.
\par However, in practice, there is a whole range of models where it remains manageable, for instance by constructing an upper bound with linear programming \cite{Huang2016}. Moreover, the tensor network formulation typically helps to deal with this question. Indeed, if the tiles $G^{\{\alpha\}}_u$ cannot tile the plane, then the associated partition function is zero in the thermodynamic limit. Thus, reciprocally, if the (exact) leading eigenvalue associated with the transfer matrix is larger than or equal to one, this implies that the plane can be tiled using $G^{\{\alpha\}}_u$ and that the lower bound is saturated. In practice, tensor network algorithms compute the contraction of the partition function and the leading eigenvalue approximately. The convergence parameter is the bond dimension of the candidate leading eigenvector in the form of a MPS. We deduce from the above that if, for increasing MPS bond dimensions, the approximate contractions converge consistently to a leading eigenvalue which is larger than or equal to one, we have numerical evidence that the maximal lower bound is saturated, implying in turn that we found the ground state manifold and its degeneracy. Conversely, if the contraction does not converge, we cannot conclude: this could either mean that the lower bound is not saturated, or that it is saturated but unavoidable spurious tiles spoil the convergence. In this case, one should try a different cluster. 

\noindent\textbf{\emph{Convergence at finite temperature}}. %
Before moving to the example, we note additionally that the above construction can in principle readily be generalised to a finite temperature tensor network. It will be the topic of another paper to show that this more generic tensor network provides accurate results and allows one to study challenging cases. Here, we just give the idea for the construction and show that it converges at finite temperatures for the triangular lattice Ising antiferromagnet.
\par The ground state tensor network that we built can be seen as the zero temperature limit of a slightly more general construction, where the $\delta$ tensor is promoted to a tensor $D_0$ (with a larger bond dimension) describing each configuration and its Boltzmann weight relative to the ground state. In the triangular lattice case, 
the finite temperature tensor network formulation of the regularised partition function $\mathcal{Z}_0$ associated with a Hamiltonian tessellation using $u$ and weights $\alpha_1 = \alpha_2 =: \alpha \in ]0,1[$ on the triangular lattice Ising antiferromagnet is thus given by bond matrices $P$ of bond dimension 16, and by $D_0$ tensors of the same bond dimension defined on each couple of triangles as
\begin{equation}
\begin{split}
	&D_0^{\{\sigma\},\{\sigma'\},\{\sigma''\},\{\sigma'''\}}(\alpha, \beta)\\
	&=\delta^{\{\sigma\},\{\sigma'\},\{\sigma''\},\{\sigma'''\}} B(\{\sigma\},\alpha, \beta)
\end{split}
\end{equation}
where the Boltzmann weight is given by
\begin{equation}
   B(\{\sigma\},\alpha, \beta) = e^{-\beta J [\alpha(\sigma_1 \sigma_2 + \sigma_1 \sigma_4) + (1-\alpha)(\sigma_2 \sigma_3 + \sigma_3 \sigma_4) + 2]}
\end{equation}
and where for short we denoted by $\{\sigma\}$ the configuration of the four spins. After SVD and grouping, the bond dimension of the network is reduced to 4.
Importantly, in the standard construction, tensor network algorithms fail to converge even at modest inverse temperatures, when $\beta$ is still small enough that the values in $t_0$ are well defined, and no ``NaN'' arise; the criterion for convergence is simply never met. In contrast, our tensor network can be contracted without issues at any inverse temperature (Fig. \ref{fig:vumps}). The zero-temperature limit of $D_0$ is well-defined by construction, and in that limit it reduces to a $\delta$ tensor corresponding to the 10 ground state tiles on $u$.  

\begin{figure}
	\centering
	\includegraphics[width=0.99\columnwidth]{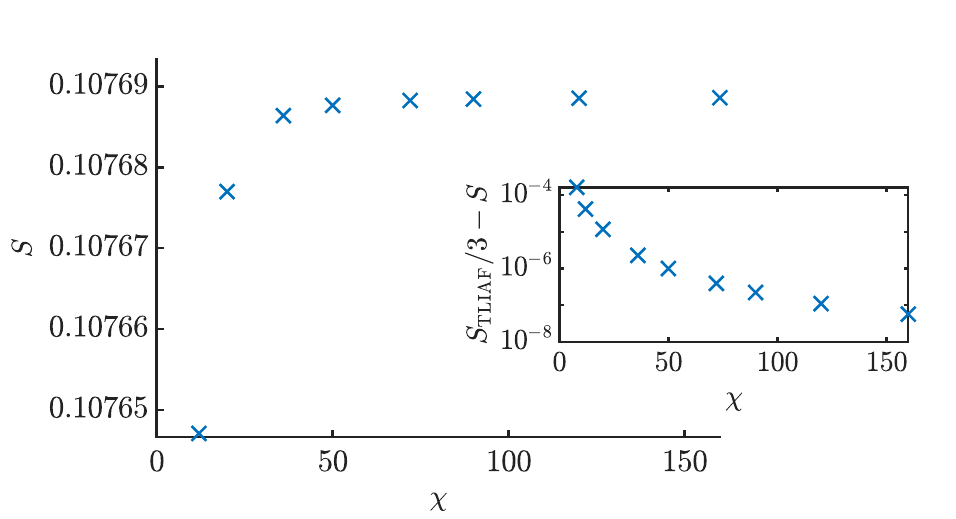}
	\caption{The residual entropy per site of the minimally frustrated tensor network for the model in Eq.~\eqref{hamiltonian}, obtained with the vumps algorithm with different bond dimensions $\chi$. In the inset we show that the value converges to a third of the value for the Ising antiferromagnet on the triangular lattice \cite{Wannier1973}.}
	\label{fig:entropy}
\end{figure}
\section{Further neighbour Ising model on the kagome lattice}
As a challenging test case, we consider a frustrated Ising model inspired by Refs.~\onlinecite{Takagi1993,Wolf1988,mizoguchi2017,Chioar16,hamp2018} and defined on the kagome lattice:
\begin{figure}
	\centering
	\includegraphics[page=18,width=0.4\columnwidth]{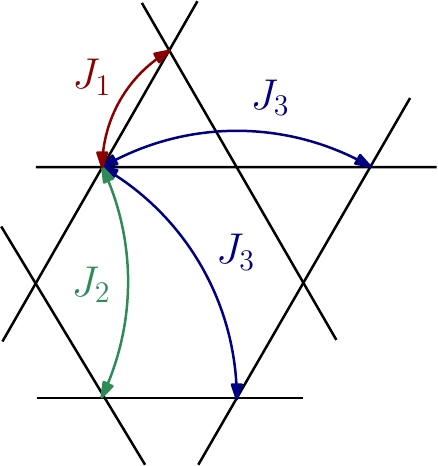}
	\caption{The two body Ising interactions present in the test model \ref{hamiltonian}.}
	\label{fig:J1J2J3}
\end{figure}
\begin{equation}
\mathcal{H} = J_1\sum_{\langle ij \rangle}\sigma_i\sigma_j + J_2\sum_{\langle\langle ij \rangle\rangle}\sigma_i\sigma_j + J_3 \sum_{\langle\langle\langle ij \rangle\rangle\rangle}\sigma_i\sigma_j\,,
\label{hamiltonian}
\end{equation}
where the sums run over (distance-based) first, second, and third nearest neighbours respectively, as is illustrated in Fig.~\ref{fig:J1J2J3}. We take $J_1=-1$ (ferromagnetic), and $J_2=J_3=10$ (antiferromagnetic). As our reference cluster $u$ for the tessellation, we use a full kagome star (12 spins, Fig.~\ref{fig:cluster}), for which 18 weights need to be determined. From the linear program, we find a ground state energy lower bound $E = \frac{2}{3} J_1 - \frac{2}{3} J_2 - J_3$ and $132$ candidate ground state tiles. The tensor network we construct for the ground state ensemble, assuming those candidate tiles, has bond dimension $18$ (very small compared to the total number of tiles of the cluster, $2^{12}=4096$). The vumps algorithm \cite{ZaunerStauber2018, Fishman2018, Vanderstraeten2019} converges nicely for this MPO for all bond dimensions of the MPS and finds a leading eigenvalue that is both real and larger than one. We thus obtain with a good level of confidence that the ground state tiles can tile the plane, which implies minimal frustration. The contraction -- which took around two hours on a laptop for the largest MPS bond dimension -- readily provides the ground state entropy to a very high precision (Fig.~\ref{fig:entropy}). Note that this method did not rely on constructing a periodic ground state, or any insight from the Monte Carlo results which are presented below; the mere existence of the state in Fig.~\ref{fig:domainwall} however \emph{proves} this result. 
\begin{figure}
	\centering
	\includegraphics[width=0.7\columnwidth]{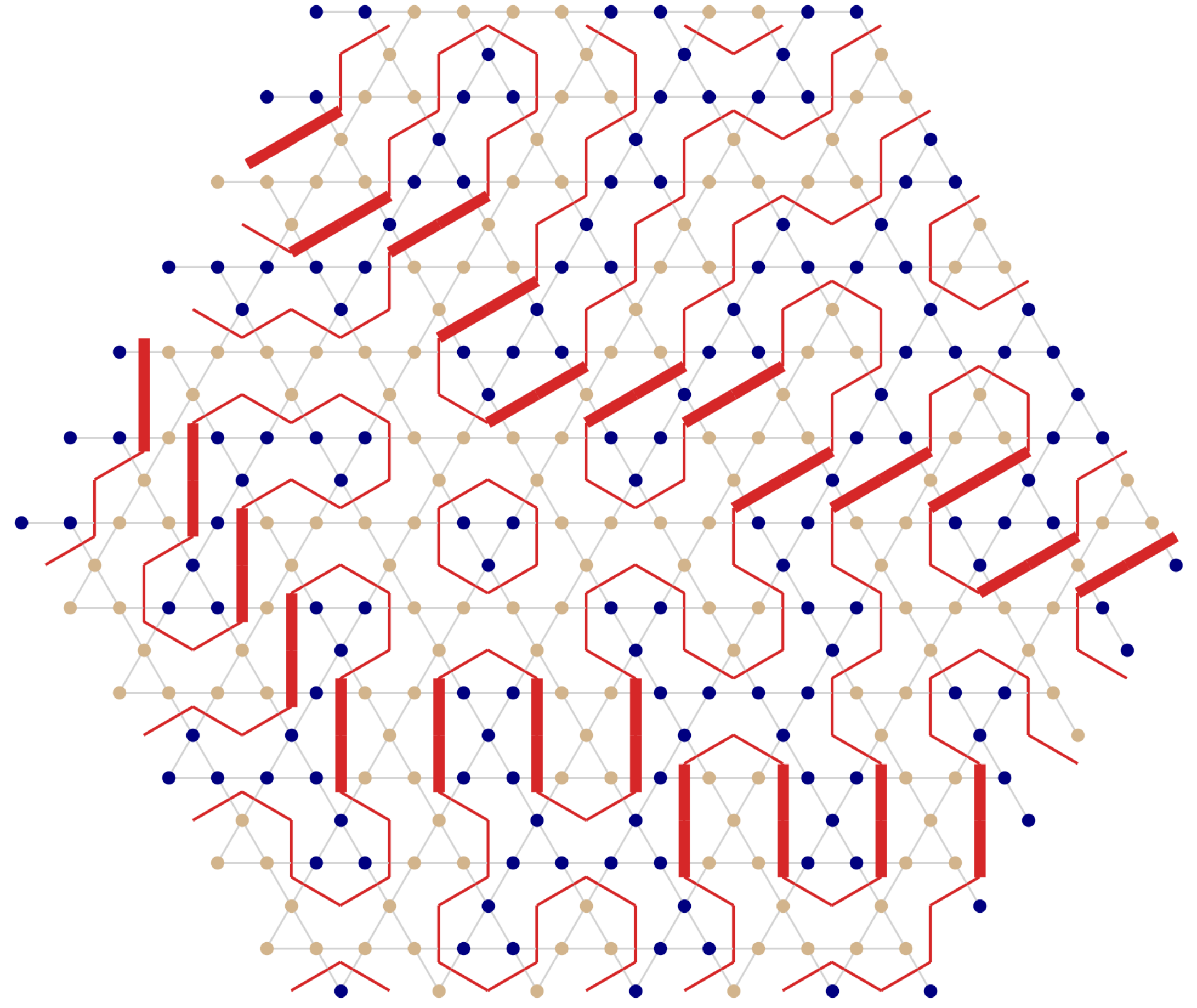}
	\caption{An example of a spin configuration in the ground state. The red lines separate up and down spins, and the red lines that cross a hexagon in a straight line have been accentuated. The tiles with a thick line separate symmetry-broken sectors where all up (resp. down) triangles are ferromagnetic. This configuration was generated during our Monte Carlo sampling and illustrates the results obtained from the tensor network.}
	\label{fig:domainwall}
\end{figure}
\begin{figure}
	\centering
	\includegraphics[width=\columnwidth]{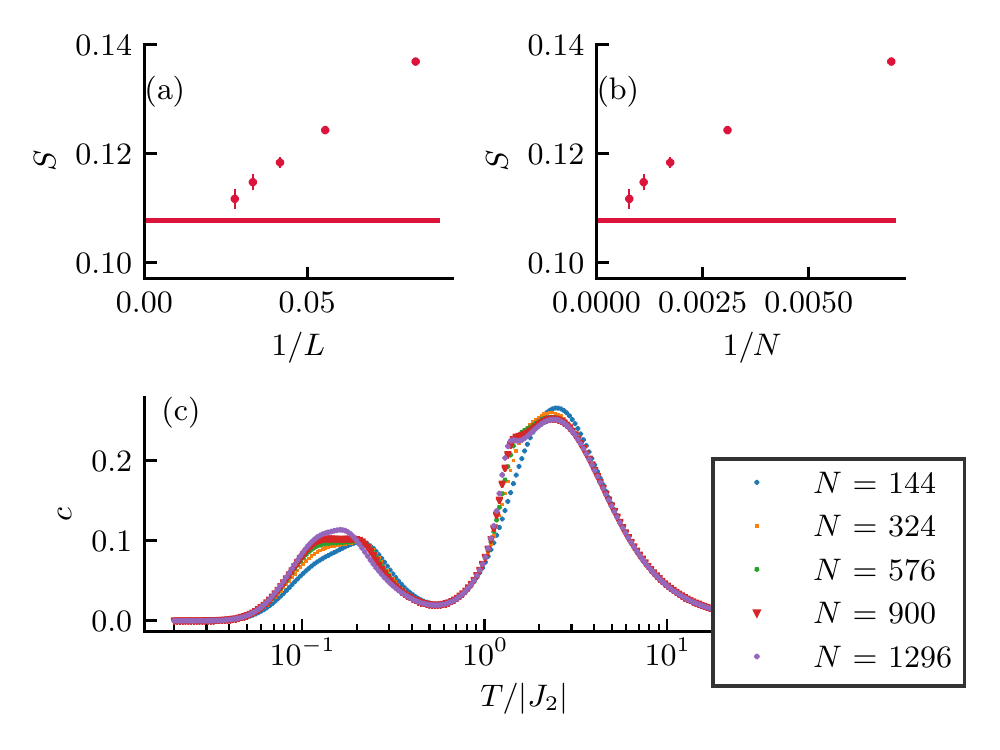}
	\caption{Monte Carlo results for the residual entropy as a function of the inverse linear system size (a) and of the inverse system size (b). For each size, the entropy is obtained by integrating the heat capacity (c) over the temperature on the whole temperature range. The heat capacity is measured on 216 temperatures thermalized with 16'384 Monte Carlo steps (MCS, consisting of 2 full updates of the state with single spin flip, 2 with the dual worm and one parallel tempering step)  and measured over 1'048'576 MCS.}
	\label{fig:MCfail}
\end{figure}
\par For comparison, we calculated the residual entropy using Monte Carlo methods (the technical details can be found in Appendix~\ref{sec:MC}). It turned out to be crucial to employ a combination of worm updates \cite{Rakala2017}, single spin flip and parallel tempering. Though one can easily generate some ground state configurations of the model (Fig.~\ref{fig:domainwall}), the evaluation of the residual entropy via thermodynamic integration is a huge challenge which requires thousands of CPU hours for a significantly less accurate result (compare Fig.~\ref{fig:MCfail} to Fig.~\ref{fig:entropy}).
\par
\begin{figure}
	\centering
	\subfigure{\includegraphics[page=16,height=0.3\columnwidth]{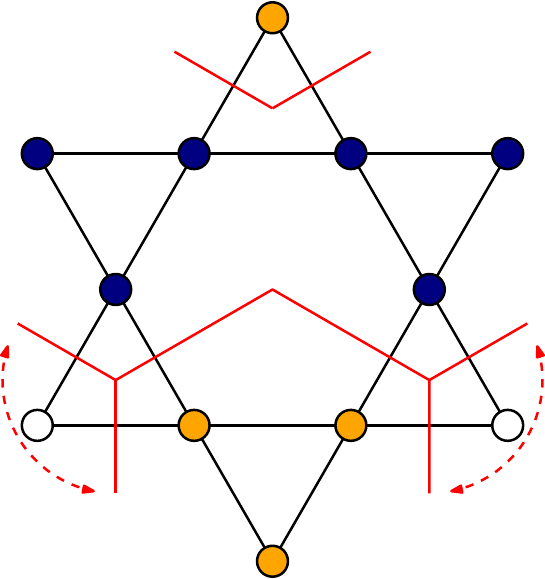}} \hspace{1cm}
	\subfigure{\includegraphics[page=17,height=0.3\columnwidth]{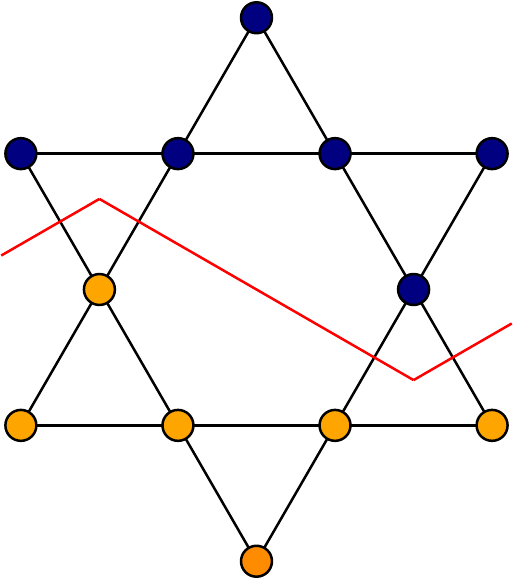}}
	\caption{Type-I tiles (left), and a type-II tile (right), with the line mapping. The arrows indicate that the line can lie across any of the two directions, but not both. The $J_2$ and $J_3$ interactions are at minimal energy, and the $J_1$ interaction is ``frustrated''.}
	\label{fig:tile}
\end{figure}

 The residual entropy obtained by contracting the tensor network is within $10^{-7}$ of one third of the triangular lattice Ising antiferromagnet entropy, suggesting some kind of correspondence between the dominant part of the ground state manifolds of both models. This correspondence can be understood thanks to exact statements based on the tiles and the tensor network construction. The 132 tiles can be split up into two types: type-I tiles for which all up (respectively down) triangles are ferromagnetic, and type-II tiles which have one up and one down antiferromagnetic triangle (Fig. \ref{fig:tile}).
To simplify the visualization, we introduce lines to separate up spins from down spins -- this line must cross the hexagon with a $120^\circ$ angle or be straight, corresponding to the two types of tiles. Furthermore, using exact tensor contractions, we find that type-I tiles form reflection-symmetry broken sectors (either up or down triangles are ferromagnetic), whereas type-II tiles (line straight across the hexagon) must appear in strings that form domain walls crossing the entire system between different reflection-symmetry broken sectors. This is illustrated by a Monte Carlo sample in Fig.~\ref{fig:domainwall}. In the ensemble of type-I tiles the ferromagnetic up, respectively down, triangles form effective degrees of freedom of a triangular Ising antiferromagnet. The residual entropy solely due to the type-I tiles is thus $\frac{1}{3}S_{TLIAF}$. 
It would thus seem that the domain walls of type-II tiles don't contribute to the residual entropy. To corroborate this we calculated the probability of finding a type-II tile on some site using the contracted tensor network and found this was zero for all bond dimensions.

\section{Outlook}

In this paper, we have introduced a general method to build contractible tensor networks for arbitrary frustrated Ising models. It relies on the identification of clusters on which the energy can be minimized independently, and on a formulation of the partition function in terms of effective degrees of freedom that correspond to all the relevant ground states on each cluster. The construction is actually possible for any model with a discrete degree of freedom, for instance Potts or clock models, and in any dimension.

To put this result in perspective, let us come back to the core of the problem faced by tensor networks for frustrated systems, namely the difficulty in numerically contracting tensors with simultaneously very large and very small elements at low temperature because of their exponential dependence on the inverse temperature $\beta$ with both positive and negative energies. This difficulty is very reminiscent of the sign problem in Quantum Monte Carlo, which excludes the investigation of the low temperature properties of a quantum system if the off-diagonal matrix elements of the Hamiltonian are not all non-positive, but which at the same time is basis dependent and can in principle be eliminated by a change of basis. What we have proposed here is a reformulation of the partition function in a basis where all elements are of the from $e^{-\beta E}$, $E\geq 0$, leading to a tensor of relatively modest dimension with only elements equal to 0 or 1 at zero-temperature, and to a contractible tensor with elements only in the interval $[0,1]$ at any positive temperature. As for the sign problem in Quantum Monte Carlo the identification of the basis does not have in general a polynomial solution. Yet, we have shown that this is in practice possible, and we will give further examples in up-coming publications, where we will study finite-temperature properties and phase transitions in highly frustrated Ising systems.

The relation to, or combination with the tropical tensor network approach to frustrated systems \cite{tropicalTN} would also be interesting and likely very fruitful.

%discussed a generic framework for studying frustrated spin systems, identifying effective degrees of freedom in these models at low temperature and enabling a tensor-network based approach. The concept of minimal frustration, and the tensor network formulation work in all dimensions.
%\par Although not the main point of this paper, the issue of contracting the standard tensor network construction for frustrated systems at relatively small inverse temperatures should not be neglected. The problem of contracting a tensor network with very large and very small numbers is  reminiscent of the infamous sign problem in Monte Carlo. It seems that this issue can pop up in any tensor network, in particular in PEPS. Here, we solved an avatar of this ``sign problem for tensor networks'', potentially showing the way to a more general solution.
\par Finally, we can consider the effect of quantum dynamics on the correlated phase spaces. Indeed we can write down PEPS wavefunctions by promoting the tiles to quantum degrees of freedom to effectively describe quantum corrections that would be present in any real life material.

\par\noindent\textbf{\emph{Acknowledgements.}} This work was initiated at the Centro de Ciencias de Benasque Pedro Pascual. FV thanks Michael Lawler for discussions about tensor networks and frustrated systems. This research is supported by ERC grant QUTE (647905) and FWO (G0E1820N) (LV, BV, FV), and the Swiss National Science Foundation (JC, FM). The Monte Carlo computations have been performed using the facilities of the Scientific IT and Application
Support Center of EPFL.

\appendix
\section{Pseudocode}
\label{sec:pseudocode}
Alg.~\ref{alg:simplex} describes a method to find a simplex of maximal dimension that fits inside a convex set $A$. The idea is to take a small simplex inside $A$ and try to make it bigger until it has the dimension of $A$. For this, the origin is first moved to the interior of the small simplex, and a vector orthogonal to the current simplex is constructed. One then looks for a vector in $A$ of maximal overlap (in absolute value) with this vector. If the maximal overlap is $0$ the simplex is of maximal dimension inside $A$, if not, one adds the result to the simplex and starts over.
\begin{figure}[H]
	\begin{algorithm}[H]
		\begin{algorithmic}[1]
			\State $\vec{R}\gets$\, random vector
			\State $\vec{\alpha}_1$\,$\gets$\, $\max \,\vec{R} \cdot\vec{\alpha}\quad$ with $\alpha\in A$
			\While{}
			\State $\vec{\beta}$ $\gets$ a point in simplex$[\{\vec{\alpha}_i\}]$
			\State Translate $\vec{\alpha}$-space by $-\vec{\beta}$
			\State $\{\vec{w}_i\}$ $\gets$ a basis of orthogonal vectors to all $\{\vec{\alpha}_i\}$
			\For{$\vec{v}\in \{\vec{w}_i,-\vec{w}_i\}$}
			\State $\vec{\alpha}$\,$\gets$\, $\max \,\vec{v} \cdot\vec{\alpha}\quad$ with $\alpha\in A$
			\If{$\vec{v} \cdot\vec{\alpha}\not=0$}
			\State Add $\vec{\alpha}$ to the set $\{\vec{\alpha}_i\}$
			\State Return to the top of the while loop
			\EndIf
			\EndFor
			\State Stop the while loop
			\EndWhile
			\State\Return $\{\vec{\alpha}_i\}$
		\end{algorithmic}
		\caption{Build interior simplex of convex set $A$}
		\label{alg:simplex}
	\end{algorithm}
\end{figure}

Alg.~\ref{alg:simplex} finds the interior simplex once the set $A$ that solves the problem
\begin{equation}
E_u\quad\leftarrow\quad\max_{\vec{\alpha}} \,E\,, \quad \text{with} \quad \left\{ \begin{array}{l} 
H^{\{\alpha\}}_u(C)\geq E \; \forall C \\ \sum_{c \in \mathcal{T}_u| n \in c} \alpha^{c}_n = 1 \end{array} \right\},
\label{linprogE}
\end{equation}
has been found. For large clusters, where there is a large number of configurations, even finding this cluster can pose memory issues. Alg.~\ref{alg:lean} offers a solution to this problem which automatically finds an interior simplex of~\ref{linprogE} using as few configurations $C$ as possible. The algorithm turns the memory load into a time load.\\
The idea is to take a restricted set of configurations $\{c_i\}$ and solve Eq. \ref{linprogE} just for them, to estimate the full solution. This amounts to finding a convex set that contains $A$. We then look for an interior simplex of the solution-set of this problem. Solving equation~\ref{linprogE}, we get a temporary estimate for the ground state energy. In each corner of the simplex, the $\vec{\alpha}$ define a Hamiltonian associating an energy to all the configurations of the clusters. The estimate is compared to the energy of each configuration in each corner. If we find a configuration which has an energy below the estimate, the associated inequality is useful; the configuration is added to the set $\{c_i\}$ and we restart. On the other hand, if in each corner, the energy of each configuration is greater than or equal to the estimate, we are sure to have a solution of Eq. \ref{linprogE} with all configurations considered, and the problem is solved. This way, we work around most of the redundancy in the set of inequalities associated with all the configurations, and only inequalities that bring insight are used to build the convex set $A$.

\begin{figure}[H]
	\begin{algorithm}[H]
		\begin{algorithmic}[1]
			\State $\{c_i\}$ $\gets$ choose some random configurations\\
			\State \emph{Add random configurations to $\{c_i\}$ until there \\ \quad is a finite $E$ and a finite interior simplex}
			%\While{}
			%\State $E_{\text{temp}}\gets$ solve Eq. \ref{linprogE} for configurations $\{c_i\}$
			%\If{$E=\infty$}
			%\State Add a random configuration to $\{c_i\}$
			%\State Go back to the top of the while loop
			%\EndIf
			%\State $\{\vec{\alpha}_i\}\gets$ Build interior simplex for $\{c_i\}$
			%\If{one of the $\vec{\alpha}_i$ lies at $\infty$}
			%\State Add a random configuration to $\{c_i\}$
			%\State Go back to the top of the while loop
			%\EndIf
			%\EndWhile
			\State\,
			\While{}
			\For{$\vec{\alpha}\in \{\vec{\alpha}_i\}$}
			\For{$c\,:\,H^{\vec{\alpha}}_u(c)< E_{\text{temp}}$}
			\If{$c\not\in\{c_i\}$}
			\State Add $c$ to $\{c_i\}$
			\State $E_{\text{temp}}\gets$ solve Eq. \ref{linprogE} for configurations $\{c_i\}$
			\State $\{\vec{\alpha}_i\}\gets$ Update interior simplex for $\{c_i\}$
			\State Return to the top of the while loop
			\EndIf
			\EndFor
			\EndFor
			\State Stop the while loop
			\EndWhile
			\State\Return $E_{\text{temp}}$\,, $\{\vec{\alpha}_i\}$
		\end{algorithmic}
		\caption{Build interior simplex of $A^u$}
		\label{alg:lean}
	\end{algorithm}
\end{figure}
Note that Alg.~\ref{alg:simplex} can just about handle the example from the paper, but larger clusters, say two or three kagome stars could only be considered using Alg.~\ref{alg:lean}.
\\

\section{Analysis of the ground state tiles}
\label{sec:rescalc}
\par The ground state tiles can be classified in two types: $48$ type-I tiles that have three non-overlapping triangles of aligned spins (i.e. ferromagnetic triangles), and $84$ type-II that don't, see Fig.~\ref{fig:tile} where we have also drawn a line separating the up from the down spins. We proceed to understand the ground state ensemble of this model by first characterizing the type-I ensemble, and then describing how type-II tiles modify this picture.
\subsection{Type-I ensemble}

\par The type-I tiles are exactly all the configurations of $u$ for which the three ferromagnetic triangles are never all pointing in the same direction. Each ferromagnetic triangle can be seen as an Ising degree of freedom; the type-I tiles are thus all the configurations for which these three degrees of freedom are never aligned. The tiles can be separated into two sub-types by reflection symmetry: the tiles where the new Ising degrees of freedom live on up triangles, and those where they live on down triangles. A global state made of tiling uniquely type-I tiles can only be made of one of these sub-types, because the tiles in one sub-type cannot be overlapping with the tiles in the other sub-type. Therefore, the type I ensemble features reflection symmetry breaking.
\par The up (down) triangles are arranged as triangular lattice, and we have seen that the type-I tiles are \emph{all} the configurations for which the three effective Ising degrees of freedom are not all aligned. So, there are no other constraints for tiling these type-I tiles. It straightforwardly follows that the effective Ising degrees of freedom must act like the spins of an Ising antiferromagnet on the triangular lattice, a model whose residual entropy is known exactly~\cite{Wannier1973}. The residual entropy of the type-I ensemble is thus given by $S = \frac{1}{3} S_{\text{TLIAF}}$. 

\subsection{Type-II ensemble}
\begin{figure}
	\centering
	\includegraphics[width = 0.8\columnwidth]{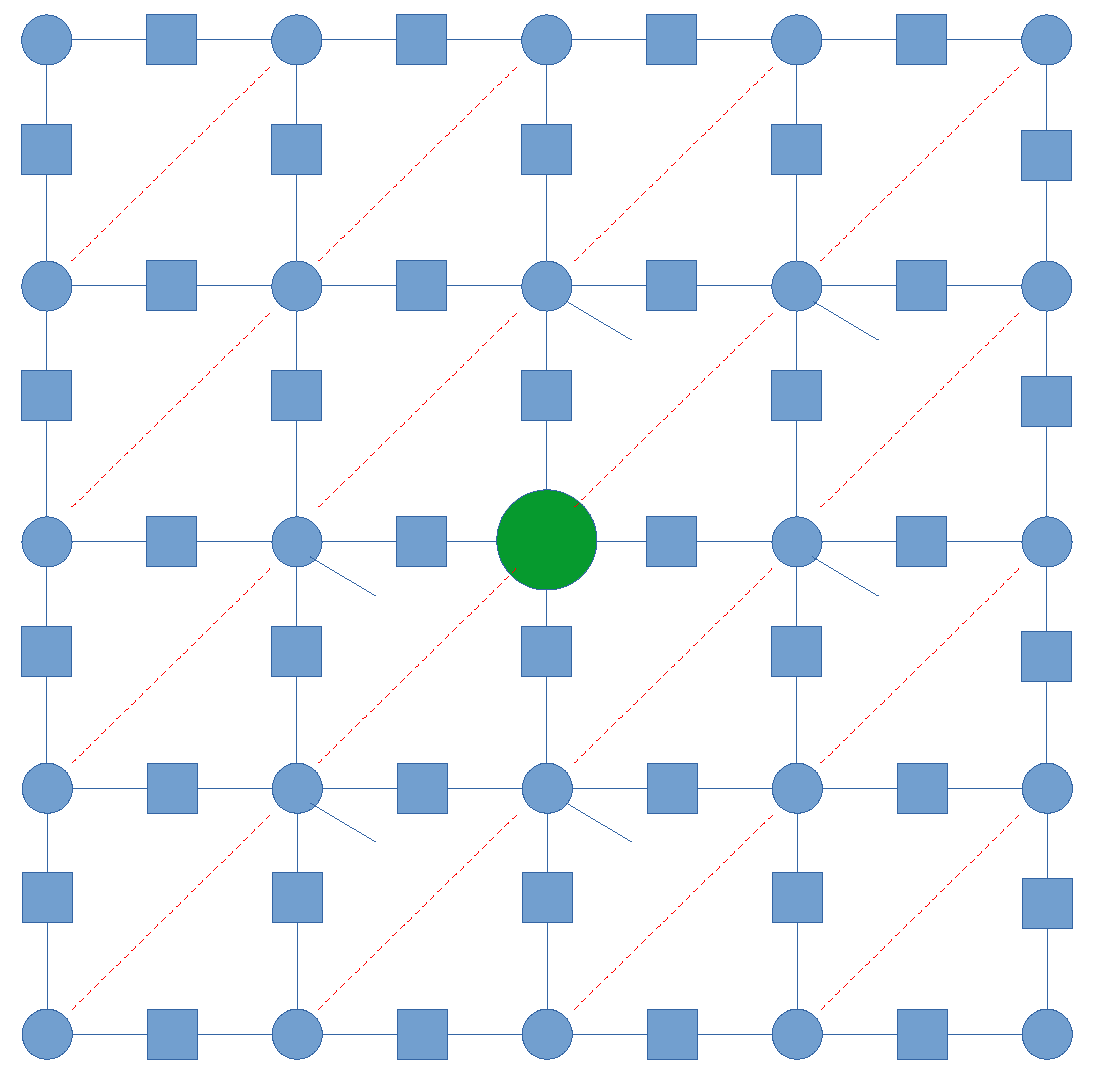}
	\caption{The vertex tensors are $\delta$ tensors representing the cluster configurations that make up the ground states, on the bonds are the usual $P$ matrices enforcing tiling rules. Of course, in practice, we perform an SVD and exact truncation of these rank-deficient matrices to be able to perform computations more efficiently. The green vertex tensor in the middle is a $\delta$ tensor that has been restricted to a single subtype of type-II tiles. The tensor network has a square lattice shape, but in reality the clusters form a triangular lattice, the red dotted lines have therefore been added to indicate nearest neighbours. To the six delta tensors describing the nearest neighbours of the central cluster we give an extra open index, allowing one to probe the local configuration.}
	\label{fig:TN}
\end{figure}
First, let us distinguish the type-II tiles from the type-I tiles. All the type-II tiles have a line (indicating an interface between up and down spins) running \emph{straight} across the hexagon. Conversely, all the configurations of the kagome star satisfying this description are type-II tiles. In type-I tiles on the other hand, the lines separating up from down spins must only live on up, respectively down triangles. An immediate consequence of this is that type-II tiles can connect type-I tiles in different reflection symmetry sectors -- if there are states containing the two types of tiles.
\par
\begin{figure}
	\centering
	\subfigure{\includegraphics[height=0.6\columnwidth]{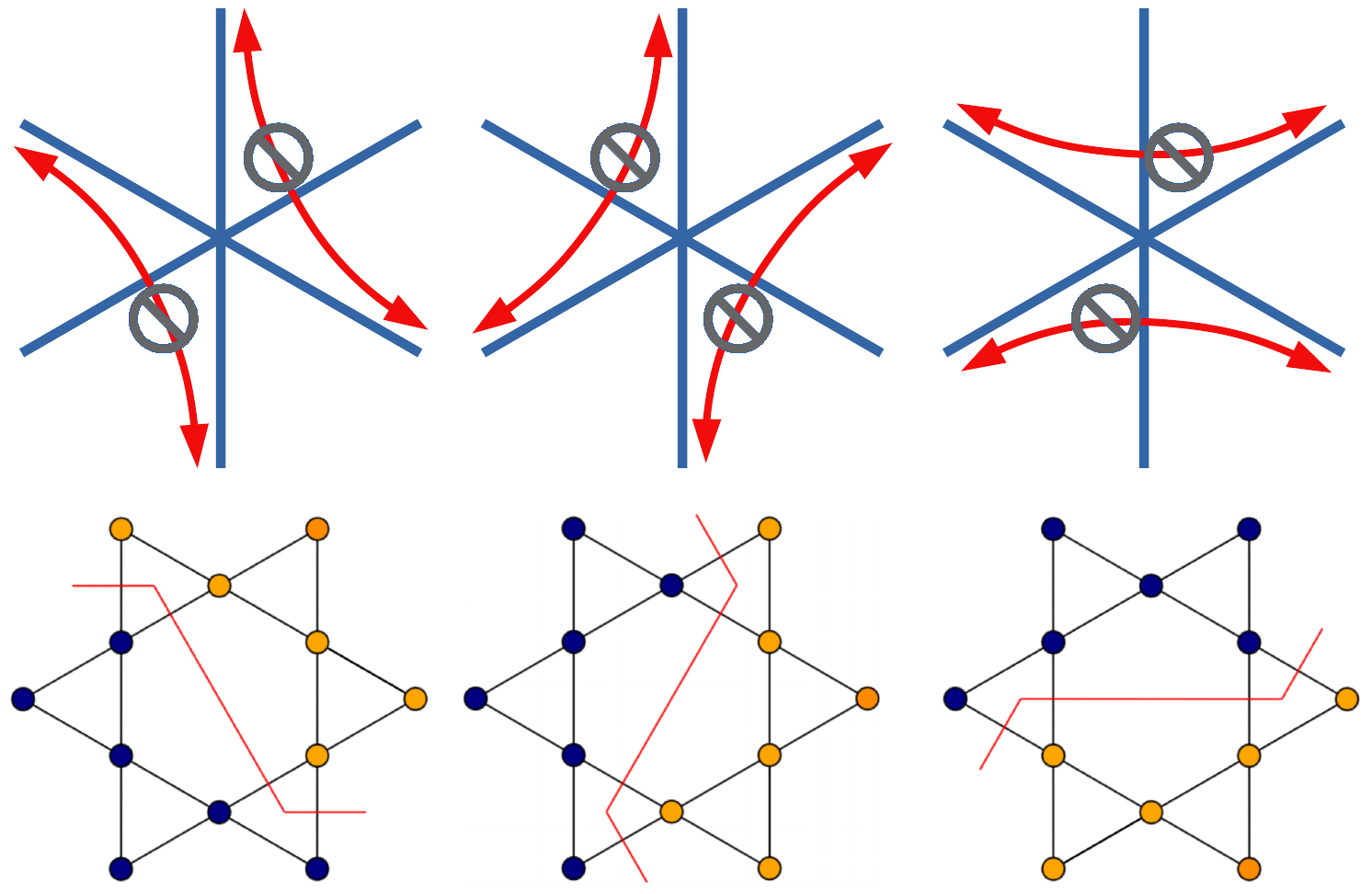}}
	\caption{The red arrows indicate which angles can not be made by the different types of domain wall. These restrictions make it such that domain walls can not make U-turns.}
	\label{fig:typeIItiles}
\end{figure}

 A key characteristic of a type-II tile is the orientation of the line crossing the hexagon. We use this to identify three subtypes of type-II tiles, illustrated in Fig.~\ref{fig:typeIItiles}. 

\par Making exact statements about how the tiles of the various subtypes can be matched together is not as easy as in the case of the type-I tiles.  To see which subtypes can neighbour one another and how, we are going to use a small tensor network construction, and exact contractions. Imagine a patch of $5\times5$ clusters where we restrict the centre tile to one subtype of the type-II tiles, and ask what types the surrounding clusters may be, while satisfying the usual tiling rules. The Tensor network for this is shown in Fig. \ref{fig:TN}. Note that this doesn't correspond to a finite system, but rather a patch of $5\times5$ in an infinite system. To each tensor neighbouring the central cluster, we add a leg, allowing one to probe the local configuration. After contraction, the indices of the resulting tensor thus correspond to labels of the tiles of the six nearest-neighbour clusters. If the value of the tensor at a certain set of indices is zero, it means that this configuration of neighbouring clusters is not allowed. Some non-zero elements may become zero if we consider a larger patch, or even only if we consider the entire infinite plane - namely, the configuration might be allowed locally but create some tiling issues at larger scales or at infinity.
\par The first result that we obtain is that a type-II tile must have \emph{exactly two} type-II tiles of the same subtype as nearest neighbours. The type-II tiles must thus make unending strings that conserve subtype, and these strings cannot cross or fuse. Additionnally, we obtain from the forbidden local cluster configurations that these strings must either go straight or make $120^\circ$ angles, but cannot make sharp angles. Moreover, of those $120^\circ$ angles, two of the six  are forbidden (which two angles depends on the subtype), making it impossible for a string of type-II tiles to close in on itself (in open boundary conditions). The forbidden angles with corresponding subtype are shown in Fig. \ref{fig:typeIItiles}
\par We thus find that the type-II tiles must form domain walls that extend the entire size of the system, separating different reflection-symmetry broken sectors made up of type-I tiles. A given domain wall can only have one specific symmetry broken sector on either side, so two neighbouring domain walls cannot be of the same type, but instead the types must alternate.
\par Finally, note that based on this analysis it is not clear whether the type-II tiles are actually tessellable, and we only found an upper bound to their freedom. We do however have a whole lot of ground states from the Monte Carlo simulations, and so we know that this picture of domain walls is indeed correct.

\section{Monte Carlo details}

\label{sec:MC}
\par For the Monte Carlo simulations, as a complement to the standard single spin flip update which is rapidly failing, we use a dual worm algorithm based on Ref.~\onlinecite{Rakala2017} as well as parallel tempering (also known as temperature replica method). For this, we use 216 walkers with a temperature associated to each walker. In a given Monte Carlo step, we first update each state with twice as many single spin flip attempts as there are sites in the system; then we perform worm updates until the total length of the worms corresponds to twice the number of dual sites of the system (see below); finally, we make a parallel tempering step. 
At each step, detailed balance is respected.

%\par In the Metropolis single spin flip algorithm, the detailed balance is respected by setting the acceptance probability to 
%\begin{equation}
%A(\left\{\sigma\right\} \rightarrow \left\{\sigma ' \right\}) = \min\left\{1, e^{-\beta \left(H\left\{\sigma'\right\}-H\left\{\sigma\right\}\right)}\right\}
%\end{equation}
\par
\begin{figure}
	\centering
	\subfigure[\label{fig:NN}]{\includegraphics[width=0.4\columnwidth]{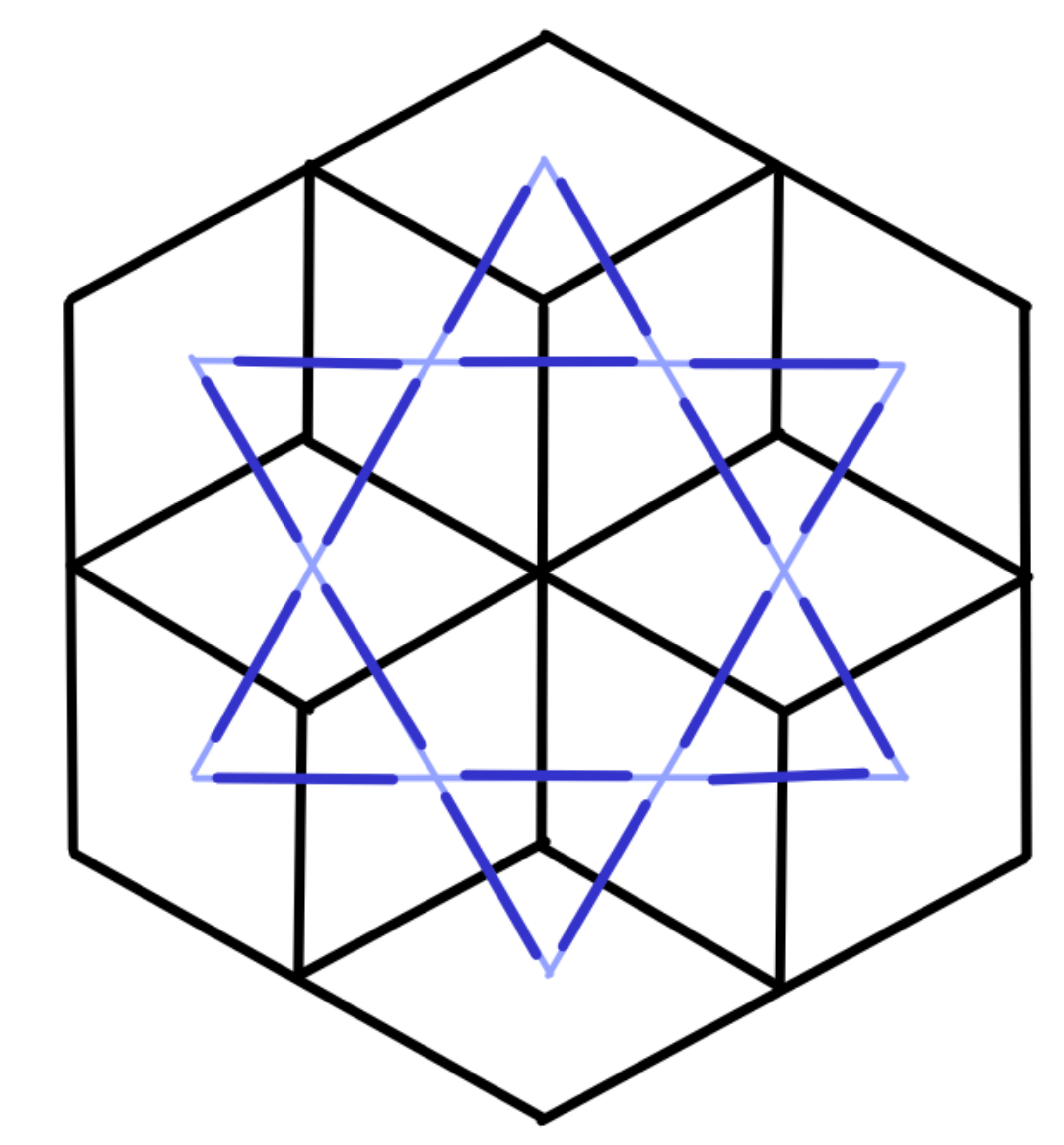}}\hfill
	\subfigure[\label{fig:2ndNN}]{\includegraphics[width=0.4\columnwidth]{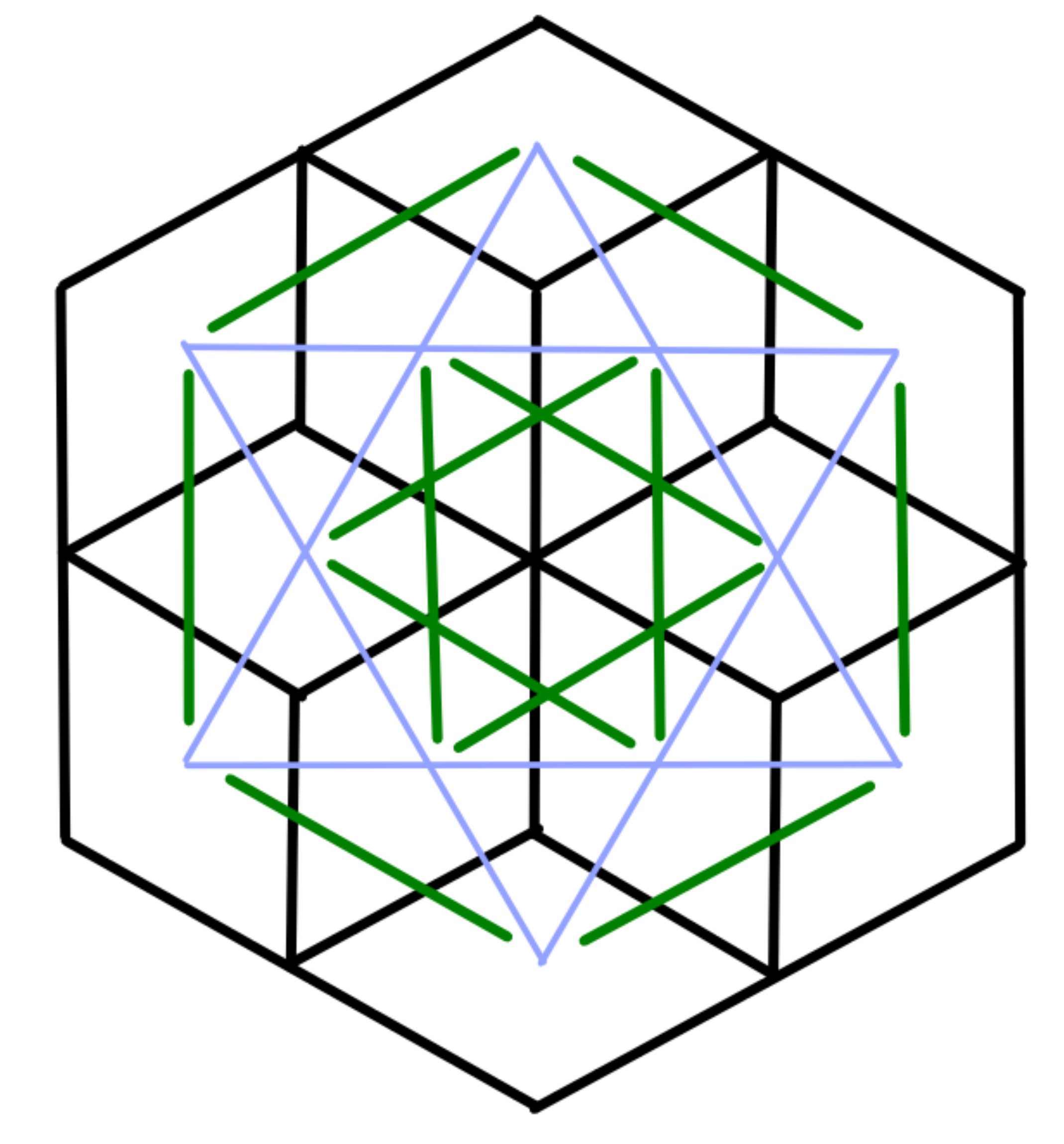}}
	\subfigure[\label{fig:3rdNNp}]{\includegraphics[width=0.4\columnwidth]{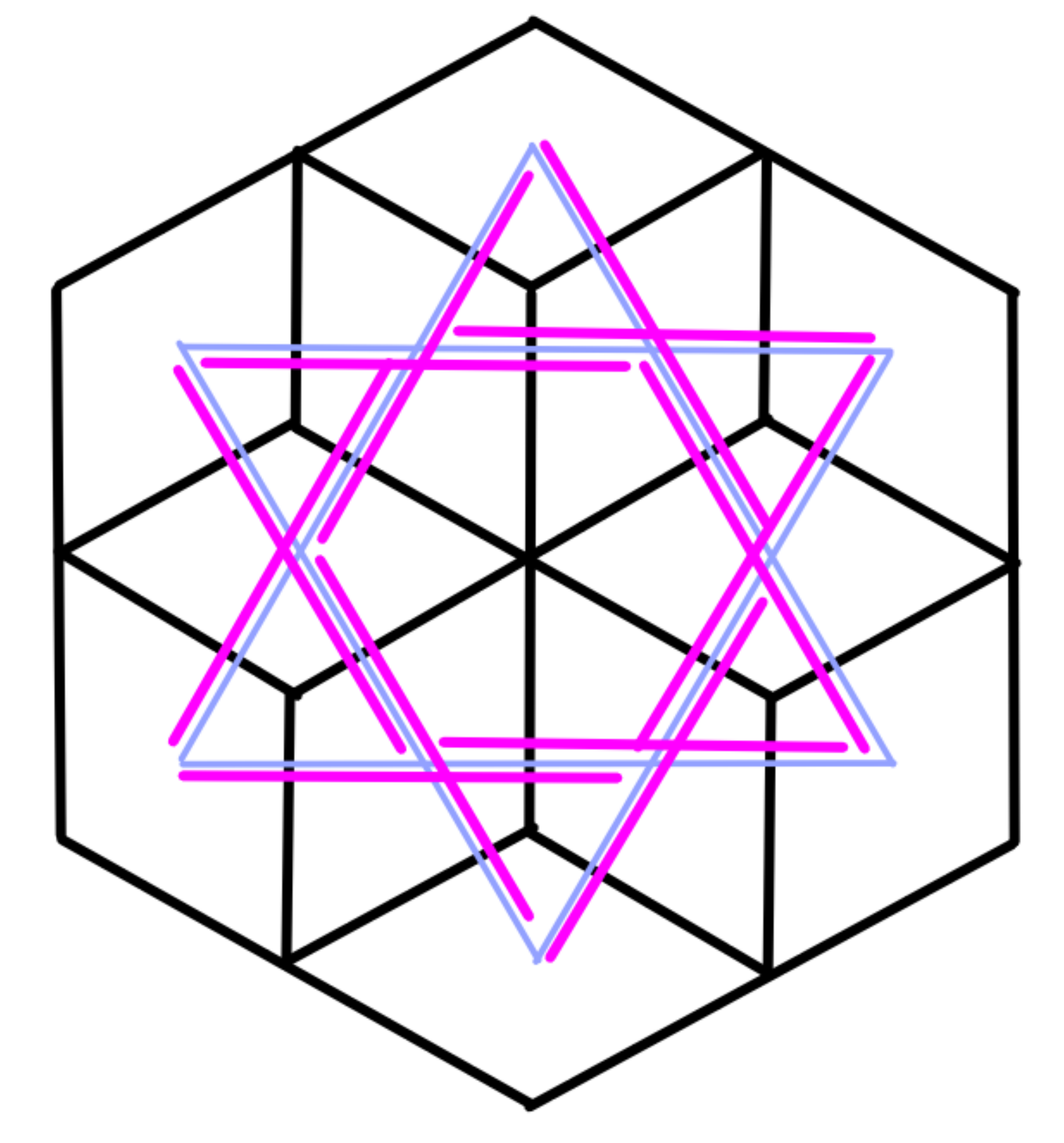}}\hfill
	\subfigure[\label{fig:3rdNNs}]{\includegraphics[width=0.4\columnwidth]{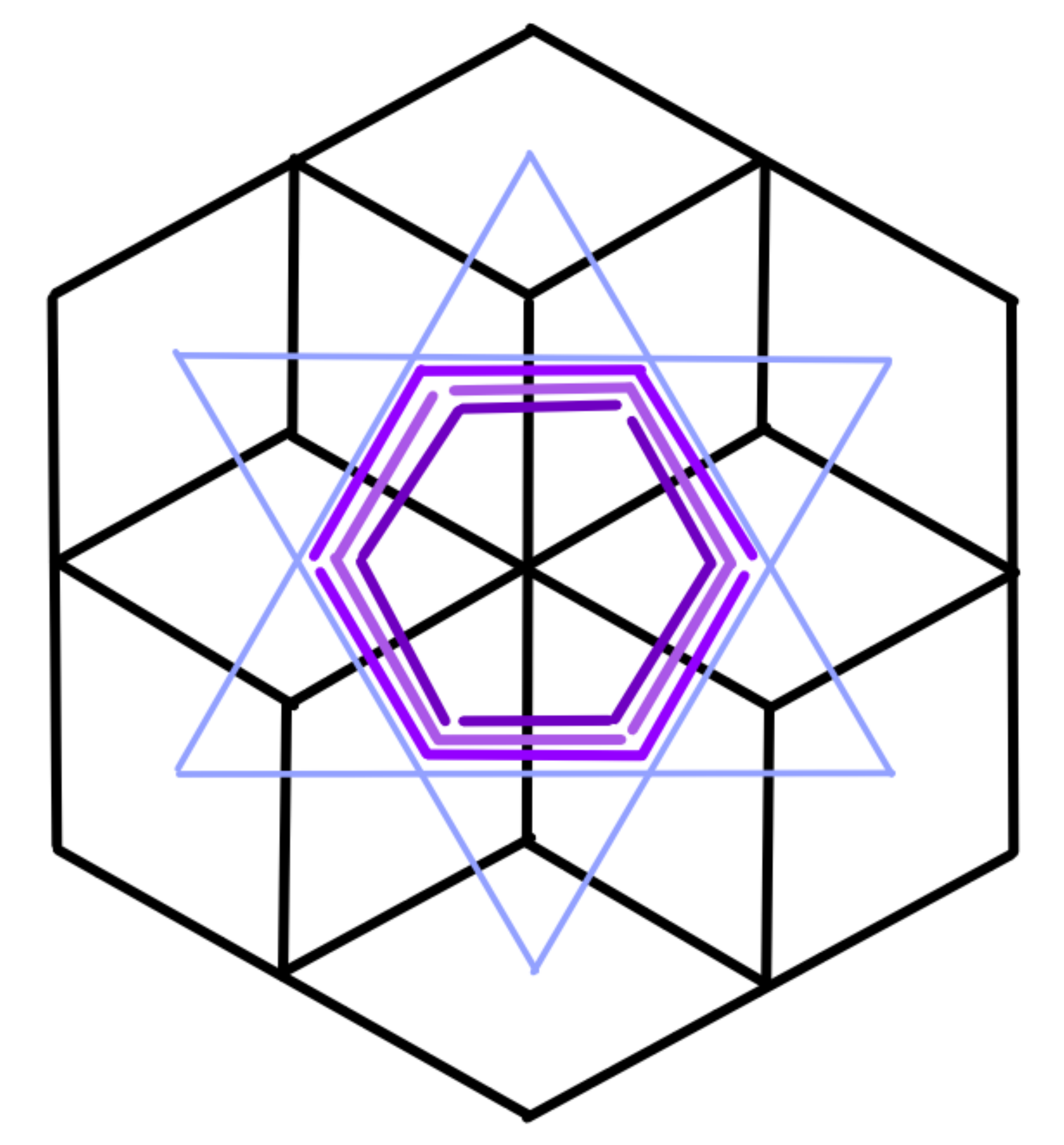}}
	\caption{Illustration of the further-neighbour interaction paths on the kagome lattice. (a) Nearest-neighbour interaction paths (coupling $J_1$). (b) Next nearest-neighbour interaction paths (coupling $J_2$). (c) 3\textsuperscript{rd} nearest-neighbour interaction paths along the bonds of the kagome lattice (coupling $J_3$). (d) 3\textsuperscript{rd} nearest-neighbour interaction paths, this time across the hexagons of the lattice (coupling $J_3$).} \label{fig:interactionpaths}
\end{figure}

 In the dual worm algorithm, the Ising model on kagome is first mapped onto a dimer model on the dice lattice according to:
\begin{align}
	H &= J_1 \sum_{\langle i,j \rangle} \sigma_i \sigma_j + J_2 \sum_{\langle\langle i,j \rangle\rangle} \sigma_i \sigma_j  + J_3 \sum_{\langle\langle\langle i,j \rangle\rangle\rangle} \sigma_i \sigma_j \\
	&= J_1 \sum_{\alpha} d_{\alpha} + J_2 \sum_{\Gamma_2} \prod_{\alpha \in \Gamma_2} d_{\alpha} + J_3 \sum_{\Gamma_3} \prod_{\alpha \in \Gamma_3} d_{\alpha}
\end{align}
where $d_{\alpha} = \sigma_i \sigma_j$ if $\alpha$ is the dual bond between the kagome lattice sites $i$ and $j$, and where $\Gamma_2$ (respectively~$\Gamma_3$) go through all the direct dimer paths connecting 2nd (3rd) nearest-neighbour spins (Fig. \ref{fig:interactionpaths}).
Building a loop update on this dimer model then corresponds to building a cluster in the original spin model. The loop is built respecting a local detailed balance condition such that, if the loop closes and the winding number of the loop in both directions on the torus is even, the update can be accepted. At the level of the dimer configuration, the local detailed balance is imposed by choosing the next direction to grow the loop uniformly at one step, and based on a weight table at the next step (see Ref.~\onlinecite{Rakala2017} and references therein for the detailed balance proof and the weight table expressions for zero-bounce and one-bounce solutions). If the winding number is odd in either direction, the updated dimer configuration on the dice lattice does \textit{not} map back to a periodic spin model on the kagome lattice. In such cases, additional loops are built until both winding numbers are even.  
\par At the parallel tempering step, it is the detailed balance of the ensemble of walkers which is respected (see for instance \cite{Katzgraber2006} and references therein). This is done by going through pairs of configurations and accepting to swap with probability
\begin{equation}
\begin{split}
&A(\left[(\{\sigma_1\}, \beta_1), (\{\sigma_2\}, \beta_2))\right] \rightarrow \left[ \beta_1 \leftrightarrow \beta_2\right]) \\
&= \min\left\{ 1, e^{( \beta_1 - \beta_2 )( H\{\sigma_1\} - H\{\sigma_2\} )}\right\}
\end{split}
\end{equation}
At even Monte Carlo steps we go through pairs starting with even indexed temperatures, while at odd steps we go through pairs starting with odd indexed temperatures.
%(\{\{\sigma_1\}, \beta_1\}, \{\{\sigma_2\}, \beta_2\})   \rightarrow (\{\{\sigma_1\}, \beta_2\}, \{\{\sigma_2\}, \beta_1\})
%1, e^{( \beta_1 - \beta_2 )( H\{\sigma_1\} - H\{\sigma_2\} )
\par These features of the Monte Carlo simulations allow one to reach the ground states. This is verified by computing the expectation value of the energy at the lowest temperature and checking that it is systematically within $10^{-9}$ of the exact ground state energies.
\par For each size, 16 independent runs are performed (for each run, 16'384 thermalization steps are followed by 1'048'576 measurement steps). 
\begin{figure}
	\centering
	\includegraphics[width = \columnwidth]{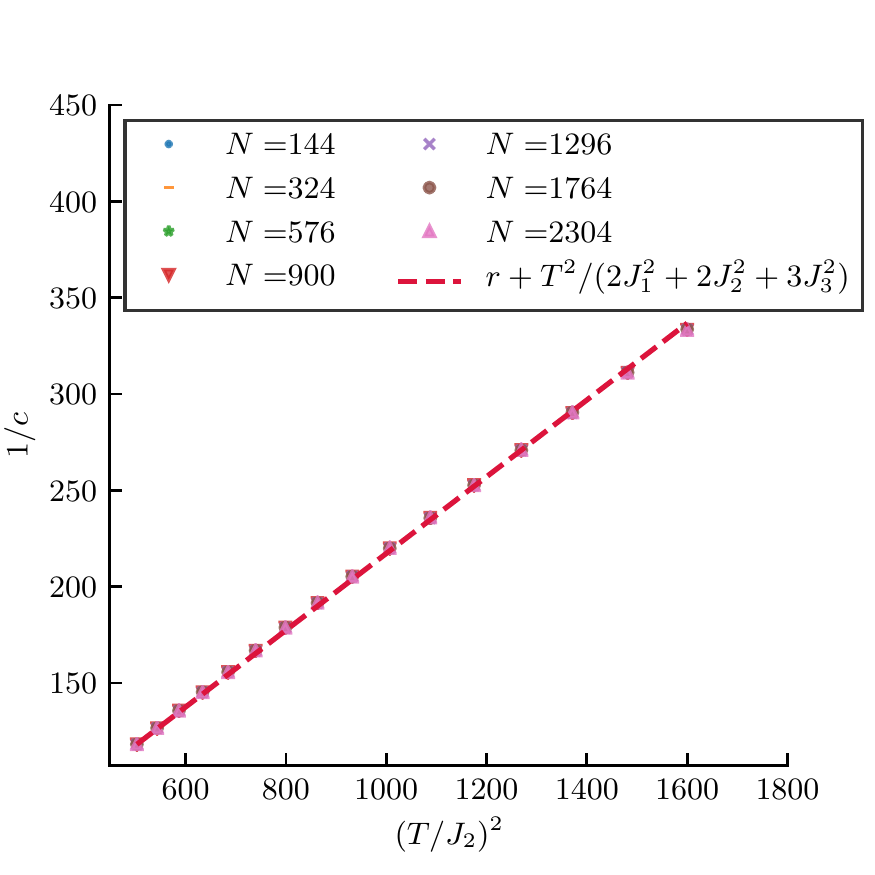}
	\caption{Comparison of the specific heat near the maximal temperature, and the high temperature expansion. The constant offset $r \cong 18 $ corresponds to a correction of order $\beta^4$ in the high temperature expansion, whose contribution to the entropy is negligible. }
	\label{fig:HighTemperatureExpansion}
\end{figure}
\begin{figure}
	\centering
	\includegraphics[width = 0.8\columnwidth]{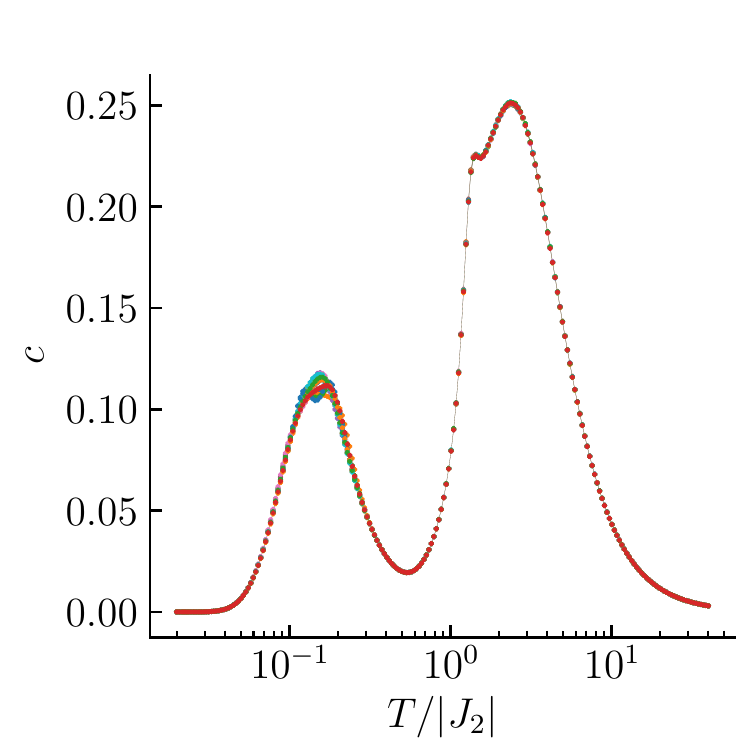}
	\caption{Specific heat for independent runs. In the lower heap, the various runs do not agree (simulations for $L = 12$).}
	\label{fig:IndepRuns}
\end{figure}
\begin{figure}
	\centering
	\includegraphics[width = 0.8\columnwidth]{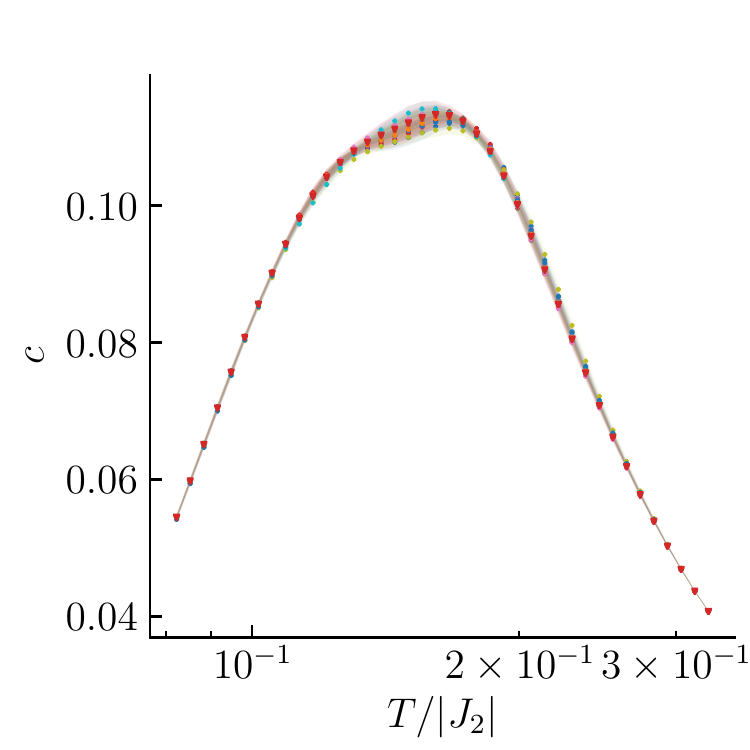}
	\caption{Performing a bootstrap analysis, the various averages of the runs agree within errorbars (zooming in on the heap at lower temperature for $L = 12$).}
	\label{fig:Bootstrap}
\end{figure}
\par The specific heat per site $c$ is computed using the variance of the energy (see for instance \cite{SandvikBible2010}) and the residual entropy per site is obtained from thermodynamic integration as:
\begin{equation}
S = \ln 2 - \int_0^\infty \frac{c}{T} \d{T}
\end{equation}
Note that in practice, we integrate numerically up to the maximal temperature $T_{\text{max}}/|J_2| = 40$. For the temperatures from $T_{\text{max}}$ to infinity, one can compute the behaviour of the specific heat from a high-temperature expansion. It is fairly easy to show that the first relevant term is 
\begin{equation}
c_{\text{High T}} = \beta^2 (2 J_1^2 + 2 J_2^2 + 3 J_3^2)
\end{equation}
and thus the integral $\int_{T_{\text{max}}}^{\infty}\frac{c}{T} \d{T} \cong 1.568 \cdot 10^{-4}$ (plotting $1/c$ as a function of $T^2$, we can check that at $T \cong T_{\text{max}}$ this first term already captures the behaviour very well, Fig. \ref{fig:HighTemperatureExpansion}). 

\par The specific heat in the lowest heap shows a dependence on the run for large sizes (Fig. \ref{fig:IndepRuns}). This is compensated for by taking the average over the 16 simulations. By a bootstrap analysis, we show that the errorbars obtained from merging the 16 independent simulations are reasonable (Fig. \ref{fig:Bootstrap}). The errorbars on the specific heat (2 standard deviations) are used to give errorbars on the residual entropy by integrating the smallest, respectively the largest possible value of the specific heat over $T$ at any temperature. Finally, we note that the reisdual entropy can alternatively be computed by integrating over the energy (see e.g. Ref.~\onlinecite{Zukovic2013} and references therein). We did this and kept only those sizes for which the simulations had been ran long enough that the two ways of computing the residual entropy would agree within errorbars.\\

\par This model seems to have extremely strong, hard to characterize finite-size effects. We show the residual entropy as a function of the inverse of $N$, the number of spins, and of $L$, the linear system size, in Fig. \ref{fig:MCfail}. Justifying finite-size corrections is a challenge, often requiring a pre-existing understanding of the ground state phase, and here we only show these two graphs as a guide to the eye. The extrapolation as a function of $1/N$ would seem to work best, at least compared to the tensor network result. However, the slope of $S(N)$  would indicate a huge prefactor to the number of ground states that we cannot explain. The extrapolation in $1/L$ would seem most plausible based on the observation that the type-II tiles form domain walls, especially since they are irrelevant to the extensive entropy. But the extrapolation in $1/L$ doesn't look too convincing and would dramatically underestimate the lower bound of $S = \frac{1}{3} S_{\text{TLIAF}}$. To solve this problem, one would need to study even larger system sizes, which turns out to be very difficult with MC, at least with our algorithm.
%Corrections in $\frac{\ln(N)}{N}$ are not unheard of and would be another possibility.  
%\par It may be that, despite rigorous checks and tests, our Monte Carlo is unable to properly sample the system. It could also be that the type-II domain walls' contribution effectively looks like a large but constant prefactor to the number of ground state for the sizes we checked. It could also be something else entirely. It is in any case clear that this model poses serious difficulty to a rigorous and precise Monte Carlo analysis. 

\bibliography{bibliography}

\end{document}